\pdfoutput=1

\documentclass[reprint,aps,twocolumn,floatfix,superscriptaddress,footinbib,prx]{revtex4-2}
\usepackage{graphicx}
\usepackage{amssymb}
\usepackage{bm}
\usepackage{amsmath}
\usepackage{amssymb}
\usepackage{amsfonts}
\usepackage{dsfont}
\usepackage{lineno}
\usepackage[english]{babel}
\usepackage[utf8]{inputenc}
\usepackage{xcolor}
\RequirePackage[T1]{fontenc}
\RequirePackage{lmodern}
\usepackage{appendix}
\usepackage{float}
\usepackage{algpseudocode}
\usepackage{algorithm}
\usepackage{multirow}
\usepackage{mathtools}
\usepackage{filecontents}
\usepackage{multibib}
\usepackage{physics}
\usepackage{comment}
\usepackage{siunitx}
\usepackage{bbm}
\usepackage{siunitx}
\usepackage{orcidlink}
\usepackage{hyperref}
\newcommand{\red}[1]{\textcolor{black}{#1}}

\begin{document}

\preprint{}
\title{Omnidirectional shuttling to avoid valley excitations in Si/SiGe quantum wells}

\author{R\'{o}bert N\'{e}meth\,\orcidlink{0009-0005-5788-3101}}
\affiliation{Department of Physics of Complex Systems, ELTE Eötvös Loránd University, H-1117 Budapest, Hungary}
\affiliation{Department of Physics, University of Wisconsin-Madison, Madison, Wisconsin 53706, USA}
\author{Vatsal K. Bandaru\,\orcidlink{0009-0008-3790-4130}}
\altaffiliation[These authors ]{contributed equally}
\affiliation{Department of Physics, University of Wisconsin-Madison, Madison, Wisconsin 53706, USA}
\author{Pedro Alves\,\orcidlink{0009-0006-0961-0405}}
\altaffiliation[These authors ]{contributed equally}
\affiliation{Department of Physics, University of Wisconsin-Madison, Madison, Wisconsin 53706, USA}
\author{Emma Brann\,\orcidlink{0009-0009-4923-1102}}
\affiliation{Department of Physics, University of Wisconsin-Madison, Madison, Wisconsin 53706, USA}
\author{Owen M. Eskandari\,\orcidlink{0009-0009-3197-9673}}
\affiliation{Department of Physics, University of Wisconsin-Madison, Madison, Wisconsin 53706, USA}
\author{Hudaiba Soomro\,\orcidlink{0009-0005-1280-510X}}
\affiliation{Department of Physics, University of Wisconsin-Madison, Madison, Wisconsin 53706, USA}
\author{Avani Vivrekar\,\orcidlink{0009-0003-6616-5625}}
\affiliation{Department of Physics, University of Wisconsin-Madison, Madison, Wisconsin 53706, USA}
\author{M. A. Eriksson\,\orcidlink{0000-0002-3130-9735}}
\affiliation{Department of Physics, University of Wisconsin-Madison, Madison, Wisconsin 53706, USA}
\author{Merritt P. Losert\,\orcidlink{0000-0002-8001-0326}}
\affiliation{Department of Physics, University of Wisconsin-Madison, Madison, Wisconsin 53706, USA}
\author{Mark Friesen\,\orcidlink{0000-0003-2878-2844}}
\altaffiliation[Contact author: ]{friesen@physics.wisc.edu}
\affiliation{Department of Physics, University of Wisconsin-Madison, Madison, Wisconsin 53706, USA}

\begin{abstract}
Conveyor-mode shuttling is a key approach for implementing intermediate-range coupling between electron-spin qubits in quantum dots.
Initial implementations are encouraging; however, long shuttling trajectories are guaranteed to encounter regions of low conduction-band valley energy splittings, due to the presence of random-alloy disorder in Si/SiGe quantum wells.
Here, we theoretically explore two schemes for avoiding valley-state excitations at these valley-splitting minima, by allowing the electrons to detour around them.
A \emph{multichannel shuttling} scheme allows electrons to tunnel between parallel channels, while a \emph{two-dimensional (2D) shuttler} provides full omnidirectional control.
Using simulations, we estimate shuttling fidelities in these two schemes, obtaining a clear preference for the 2D shuttler.
Based on such encouraging results, we propose a \red{modular qubit architecture} based on 2D shuttling, which enables all-to-all connectivity within qubit plaquettes and high-fidelity communication between different plaquettes.
\end{abstract}

\maketitle

It is anticipated that a large-scale quantum computer formed of quantum-dot spin qubits will require some type of intermediate-range quantum coupler~\cite{Vandersypen:2017p34}.
Currently, two schemes are studied most actively: the \emph{bucket-brigade} shuttler~\cite{Fujita:2017p22, Mills:2019p1063, Ginzel:2020p195418, Buonacorsi:2020p125406, Yoneda:2021p4114, Jadot:2021p570, Krzywda:2021p075439, Noiri:2022p5740, Boter:2022p024053, Noiri:2022p5740, Langrock:2023p020305, Zwerver:2023p030303, Sato:2025p1, DeSmet:2025p866}, in which electrons or holes are passed sequentially between dots in a linear array by modulating their detuning potentials, and the \emph{conveyor-mode} shuttler~\cite{Taylor:2005p177, Seidler:2022p100, Langrock:2023p020305, Ermoneit:2023preprint, Kunne:2024p4977, Struck:2024p1325, Xue:2024p2296, Losert:2024p040322, DeSmet:2025p866, Mokeev:2024arXiv, Jeon:2025p195302, Oda:2024arXiv}, in which a qubit is transported within a moving potential pocket.
Initial results are encouraging:
electron charges have been shuttled with high fidelities over distances of $\sim$$\SI{20}{\micro\meter}$~\cite{Xue:2024p2296, Seidler:2022p100, Mills:2019p1063}, phase-coherent shuttling has been demonstrated over a distance of $\SI{400}{\nano\meter}$~\cite{Struck:2024p1325}, and independent spins have been shuttled back and forth within a few-dot array over a total distance of $\SI{80}{\micro\meter}$~\cite{Zwerver:2023p030303}.
Conveyor-mode schemes are found to be economical in terms of their control lines, and recent reports indicate that they may also provide higher shuttling fidelities than the bucket brigade~\cite{Langrock:2023p020305, DeSmet:2025p866}.
In this paper, we consider the conveyor-mode approach.

\begin{figure}[t] 
	\includegraphics[width=8cm]{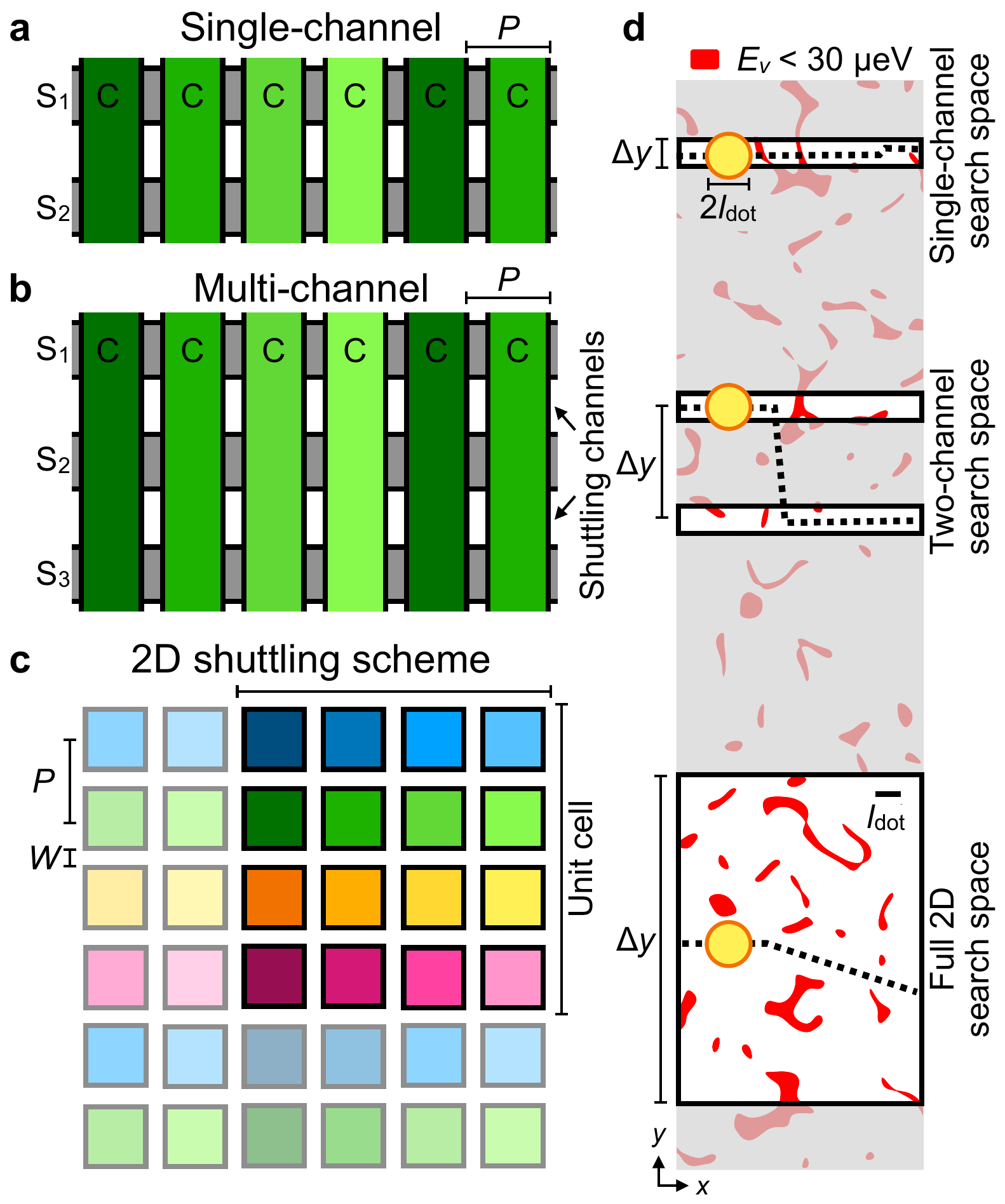}
	\centering
	\caption{
    Shuttling schemes for avoiding regions of low valley-energy splittings.
(a) A conventional \emph{single-channel} scheme is formed of one shuttling channel, surrounded by screening gates (S$_1$ and S$_2$) that provide limited control of electron motion transverse to the channel. 
Clavier gates (C) with gate pitch $P$ are formed into unit cells (indicated by shading); sinusoidally varying voltage signals provide a moving potential pocket that can transport electrons along the channel.
(b) A \emph{multichannel} scheme provides greater transverse motion by defining two or more channels.
Independent voltage control of the screening gates (S$_1$-S$_3$) allows for control of the energy detuning and tunnel coupling between the channels.
(c) A \emph{2D shuttling} scheme is defined by pixel-like ``clavette’’ gates, formed into a 2D unit cell, with gate pitch $P$ and separation $W$; sinusoidally varying voltage signals now provide omnidirectional control of the moving potential pocket.
(d) A typical map of low valley splittings, similar to those calculated in~\cite{Losert:2023p125405}.
Single-channel, two-channel, and 2D shuttling geometries provide increasing levels of transverse shift control (dotted lines), $\Delta y$, to avoid regions with low valley splittings.
Here, we assume an average valley splitting of ${\bar E}_v = 100$~\SI{}{\micro\electronvolt} and dot diameter $2l_\text{dot} = 2\sqrt{\hbar^2/m_t E_\text{orb}} \approx 28$~\SI{}{\nano\meter} (black scale bar), consistent with a lateral confinement energy of $E_\text{orb} = 2$~\SI{}{\milli\electronvolt}.
}
	\label{fig:1}
\end{figure}

The main challenge for conveyor-mode shuttling in Si/SiGe quantum wells arises from locally varying materials parameters and confinement potentials – a common problem for solid-state devices – which can cause excitations outside the computational subspace~\cite{Langrock:2023p020305}.
Such disturbances include electrical (``charge’’) fluctuations, magnetic fluctuations, and most prominently, fluctuations of the conduction-band valley-state energy splitting (the ``valley splitting’’), induced by atomistic disorder~\cite{Friesen:2006p202106, Friesen:2007p115318, Kharche:2007p092109, Culcer:2010p205315, Gamble:2013p035310, Boross:2016p035438, Abadillo-Uriel:2018p165438, Tariq:2019p125309, Hosseinkhani:2020p043180, Dodson:2022p146802} – in particular, disorder of the random SiGe alloy~\cite{Wuetz:2022p7730, Losert:2023p125405, Lima:2023p025004, Pena:2023p33}. 
In the alloy-disorder-dominated (ADD) regime, which is thought to encompass all current experiments, the average valley splitting $\bar E_v$ depends strongly on the overlap of the wave function with Ge atoms, while the standard deviation of the valley splitting is given by $\sigma_\Delta \approx \bar E_v/\sqrt{\pi}$~\cite{Losert:2023p125405}.
The shuttling electron is therefore assured of encountering sites with dangerously low valley splittings, given a long-enough shuttling trajectory~\red{\cite{Langrock:2023p020305,Losert:2024p040322}}.
At such locations, the electron is likely to suffer a harmful valley excitation through a Landau-Zener process.
It has recently been shown that the best approach for suppressing these excitations is to apply multiple strategies simultaneously ~\cite{Losert:2024p040322}.
The most important strategies include (1) modifying the quantum-well composition to increase $\bar E_v$ (e.g., by adding a small amount of Ge to the well), and (2) allowing the shuttling trajectory to be shifted transversely, to detour around the dangerous sites.
[See Fig.~\ref{fig:1}(d). Note that the latter strategy implicitly requires obtaining a 2D map of $E_v$ across the shuttler, as demonstrated in~\cite{Volmer:2024p61}.]
Here, the magnitude of the transverse shift $\Delta y$ should be somewhat larger than the dot diameter $2l_\text{dot}$; for a typical value of $2l_\text{dot}=28$~nm, it was found that $\Delta y =100$~nm is sufficient for providing good shuttling fidelities.
However, in conventional quantum dot devices~\cite{Dodson:2022p146802} and shuttlers~\cite{Volmer:2024p61}, realistic shifts are usually no larger than $\Delta y \approx 20$~nm, which is insufficient for achieving high fidelities.
Conventional shuttling devices therefore have a critical limitation: they are only designed for one-dimensional (1D) operation.
True 2D operation, needed to suppress valley excitations, requires rethinking the underlying architecture.

In this work, we propose two shuttling schemes that allow for 2D motion.
We first propose to extend the conventional 1D shuttling geometry [Fig.~\ref{fig:1}(a)] by introducing parallel shuttling channels separated by screening gates [Fig.~\ref{fig:1}(b)].
By enabling tunneling between these channels, we can achieve a significant enhancement of $\Delta y$.
There are no technological obstacles to implementing such a scheme because it uses existing overlapping-gate fabrication methods; the tunneling procedure is somewhat error-prone, however, as we demonstrate in our simulations below.
Our second proposal comprises a fully 2D architecture, obtained by tiling a 2D unit cell of ``clavette’’ gates [Fig.~\ref{fig:1}(c)], which enable conveyor-mode shuttling in arbitrary directions using only a limited number of ac signals.
These devices may be fabricated by etched deposition and vertical ``via’’ wiring methods, as demonstrated recently in industrial settings~\cite{Ha:2022p1443, Acuna:2024p044057, George:2025p793}.
Our simulations suggest that such 2D schemes can overcome many of the fidelity challenges encountered in quasi-1D shuttling.
Motivated by the flexibility of this 2D platform, we conclude by proposing a scalable quantum-dot architecture based on 2D shuttling.

\section{Multichannel shuttling}
\begin{figure*}[t] 
	\includegraphics[width=17.8cm]{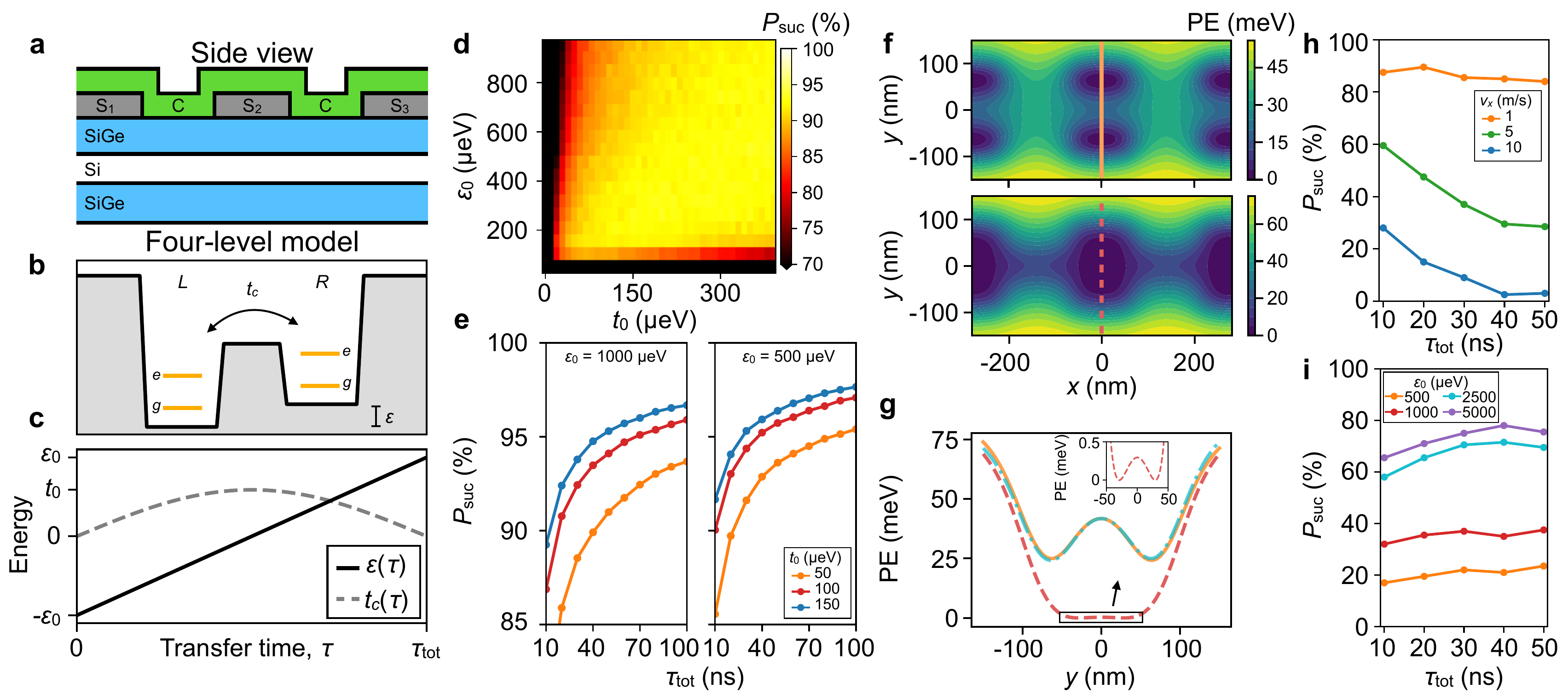}
	\centering
	\caption{
    Multichannel shuttling simulations. 
(a) A schematic side view of a shuttling device, showing a Si/SiGe quantum well with clavier (C) and screening (S) gates. 
(b) A schematic illustration of the four-level model used to simulate valley leakage while transferring the qubit from the left channel ($L$) to the right channel ($R$). 
 We include ground ($g$) and excited ($e$) valley states in each channel, and define the detuning $\varepsilon$ and tunnel coupling $t_c$ between the channels.
(c) Illustration of a typical modulation schedule for $\varepsilon(\tau)$ and $t_c(\tau)$, for performing a channel transfer. 
``Paused'' protocol (d)-(g):
(d) The computed success probability for a transfer, $P_\text{suc}$, as a function of $t_0$ and $\varepsilon_0$. 
(e) $P_\text{suc}$ as a function of the total transfer period $\tau_\text{tot}$, for the parameter values $\varepsilon_0 = 1000$~\SI{}{\micro\electronvolt} (left) and \SI{500}{\micro\electronvolt} (right). 
Each data point in (d) and (e) corresponds to an average over 10,000 instances of random-alloy disorder. 
(f) 2D potential-energy (PE) landscapes obtained at two times during the transfer process: $\tau = 0$ (top) and $\tau =  \tau_\text{tot}/2$ (bottom). 
By varying the screening gate voltages, we can tune both $\varepsilon_0$ and $t_0$; for these two simulations we obtain $\varepsilon_0 = 750$~\SI{}{\micro\electronvolt} and $t_0=200$~\SI{}{\micro\electronvolt} (see Appendix~\ref{app:electrostatics}), which give a high probability for success, as indicated in (d).
(g) 1D linecuts through the electrostatic potential-energy landscapes shown in (f). 
Here, we include cuts through $x = 0$ for the cases $\tau = 0$ (orange), $\tau = \tau_\text{tot}/2$ (red), and $\tau = \tau_\text{tot}$ (cyan).
The detuning between the cyan and orange curves is barely visible at this scale.
Inset: a blown-up view of the double-dot potential for the case $\tau = \tau_\text{tot}/2$. 
At this point, the barrier height between dots along the channel axis (along $\hat{x}$, not shown) is still $>15$~\SI{}{\milli\electronvolt}.
``Moving'' protocol (h),(i):
(h) Transfer success probabilities in the ``correlated'' disorder regime, assuming $\varepsilon_0 = 500$~\SI{}{\micro\electronvolt} and $t_0 = 100$~\SI{}{\micro\electronvolt}, for the cases $v_x = 1$~\SI{}{\meter\per\second} (orange), \SI{5}{\meter\per\second} (green), and \SI{10}{\meter\per\second} (blue).
(i) Transfer success probabilities obtained in the ``uncorrelated'' disorder regime, assuming $v_x = 1$~\SI{}{\meter\per\second} and $t_0 = 100$~\SI{}{\micro\electronvolt},  for the cases $\varepsilon_0 = 500$~\SI{}{\micro\electronvolt} (orange), 1000~\SI{}{\micro\electronvolt} (red), 2500~\SI{}{\micro\electronvolt} (cyan), and \SI{5000}{\micro\electronvolt} (purple).
Each data point in (h) and (i) is averaged over 200 simulations with randomly generated disorder.}
\label{fig:2}
\end{figure*}

We first consider the multichannel shuttler illustrated in Figs.~\ref{fig:1}(b) and \ref{fig:2}(a).
In this device, two (or more) parallel channels are separated by screening gates, with shared overlapping clavier gates that enable conveyor-mode shuttling.
This arrangement allows an electron to be shuttled within a single channel, similar to single-channel shuttling schemes~\cite{Langrock:2023p020305}.
However, the proximity of the second channel also allows for tunneling between channels, more similar to bucket-brigade operation.
Since the channels are separated by the width of a screening gate, this arrangement can provide an effective $\Delta y \gtrsim 100$~\SI{}{\nano\meter}.
The main new source of infidelity in this geometry arises from interchannel tunneling, which, like bucket-brigade shuttling, suffers from lower fidelities.

We now perform simulations to characterize the fidelity of the multichannel shuttling scheme.
We consider the four-level system illustrated in Fig.~\ref{fig:2}(b), with two channels ($L$ and $R$) and two energy levels in each channel (ground and excited, $g$ and $e$).
The potential energy difference between the channels is defined as the detuning $\varepsilon$, and transitions between the channels are enabled by the tunnel coupling $t_c$.
These Hamiltonian parameters are controlled by the voltages applied to the three screening gates [S$_1$-S$_3$, Fig.~\ref{fig:2}(a)] \red{and the clavier gates (C), where S1 and S3 have the strongest effect on $\varepsilon$, while S2 has the strongest effect on $t_c$.} 
The ground and excited levels represent valley eigenstates, where ``valleys’’ refers to the energy minima of the Si conduction-band structure, labeled $| z_\pm \rangle $, which occur at locations $\pm k_0 \hat {\bm z}= \pm 0.82 (2 \pi / a_0) \hat {\bm z}$ in the Brillouin zone, and $a_0 = 0.543$~\SI{}{\nano\meter} is the cubic lattice constant of the Si crystal unit cell~\cite{Friesen:2007p115318}.
Since spin-orbit coupling is weak in Si and SiGe, the spin states effectively decouple from the valley-orbit states over the timescales considered in this work; we therefore explore the valley-orbit physics and ignore the spin physics here.
Over these relatively short timescales, we can also safely ignore relaxation processes.

During successful shuttling, the qubit may only fill the ground state of the appropriate shuttling channel – any other level occupation represents ``leakage,’’ which contributes significantly to the infidelity of the shuttling process~\cite{Langrock:2023p020305, Losert:2024p040322}.
For an ideal transfer operation between the two channels, tunneling should be performed adiabatically, 
so the system remains in the ground state of the full 4D Hamiltonian.
However, such ideal operation is complicated by the facts that (1) the coupling between valleys within a given channel, defined as $\Delta = \langle z_+ | H | z_- \rangle$ (with the corresponding valley splitting $E_v = 2|\Delta|$), depends strongly on the local Ge concentration disorder~\cite{Losert:2023p125405}, both in its magnitude and its complex phase, and (2) tunneling is only permitted between states having the same valley index ($+$ or $-$).
Tunneling can therefore cause the ground valley state in one channel to be projected onto the excited valley state in the second channel.
Such effects are captured in the following Hamiltonian, expressed in the $\{|L,z_+\rangle,|L,z_-\rangle,|R,z_+\rangle,|R,z_-\rangle\}$ basis:
\begin{multline} \label{eq:ham_init}
H = \frac{\varepsilon}{2} \tau_z + t_c \tau_x + \mathcal{P}_L \left( \text{Re}[\Delta_L] \gamma_x - \text{Im}[\Delta_L] \gamma_y \right) \\
+ \mathcal{P}_R \left( \text{Re}[\Delta_R] \gamma_x - \text{Im}[\Delta_R] \gamma_y \right) .
\end{multline}
Here, $\Delta_{L(R)}$ is the intervalley coupling in the left (right) shuttling channel, the operators $\tau_j$ are Pauli operators acting in channel space ($\tau_z = |L \rangle \langle L | - | R \rangle \langle R |$ and $\tau_x = |R \rangle \langle L | + | L \rangle \langle R | $), $\gamma_j$ are Pauli operators acting in valley space, and $\mathcal{P}_{L(R)}$ are projection operators acting on the left (right) channel subspace ($\mathcal{P}_{L(R)} = |L(R)\rangle \langle L(R)|$).

To simplify our simulations, we now rotate Eq.~(\ref{eq:ham_init}) into a basis that locally diagonalizes the valley states.
This rotation is given by $U_v = \mathcal{P}_L U_L + \mathcal{P}_R U_R $, where $U_{L(R)} = (1/\sqrt{2})(\gamma_0 + i \gamma_y \cos \phi_{L(R)} + i \gamma_x \sin \phi_{L(R)})$, and $\phi_{L(R)} = \text{Arg}[\Delta_{L(R)}]$ is known as the valley phase.
The resulting Hamiltonian in the basis $\{|L,e\rangle,|L,g\rangle,|R,e\rangle,|R,g\rangle\}$ is given by $\tilde H = U_v H U_v^\dag$, such that
\begin{equation} \label{eq:total_ham}
\tilde H = \begin{pmatrix}
\frac{\varepsilon}{2} + |\Delta_L| & 0 & t_{ee} & t_{eg} \\
0 & \frac{\varepsilon}{2} - |\Delta_L| & t_{ge} & t_{gg} \\
t_{ee}^* & t_{ge}^* & -\frac{\varepsilon}{2} + |\Delta_R| & 0 \\
t_{eg}^* & t_{gg}^* & 0 & -\frac{\varepsilon}{2} - |\Delta_R|
\end{pmatrix} ,
\end{equation}
where the tunneling matrix elements are given by
\begin{gather}\label{eq:tunneling_elements}
t_{gg} = \frac{t_c}{2} \left( e^{-i (\phi_L - \phi_R)} + 1 \right) , \\
t_{ee} = \frac{t_c}{2} \left( e^{i (\phi_L - \phi_R)} + 1 \right) , \\
t_{eg} = \frac{t_c}{2} \left(e^{i \phi_L} - e^{i \phi_R} \right) , \label{eq:teg} \\
t_{ge} = \frac{t_c}{2} \left(e^{-i \phi_R} - e^{-i \phi_L} \right)  . \label{eq:tge}
\end{gather}

Equations~(\ref{eq:total_ham})-(\ref{eq:tge}) form the starting point for our shuttling simulations.
In the ADD regime, $\Delta$ fluctuates strongly, due to local random-alloy disorder.
While correlations in $\Delta$ values occur between dots separated by short distances~\cite{Losert:2023p125405}, the 100~nm channel separation assumed in this work effectively suppresses such correlations~\cite{Marcks:2025preprint}.
Consequently, $\Delta_L$ and $\Delta_R$ are taken to be uncorrelated, with values drawn from a complex normal distribution, with zero mean and a variance given by $\sigma_\Delta$ \cite{Wuetz:2022p7730, Losert:2023p125405}.
In particular, the valley phases in the channels are independently randomized, leading to nonzero valley phase differences, $\delta \phi = \phi_L - \phi_R$.
From Eqs.~(\ref{eq:teg}) and (\ref{eq:tge}), we see that $\delta \phi\neq 0$ leads to nonzero intervalley couplings, $t_{eg}$ and $t_{ge}$, which can induce valley excitations as the dot is transferred between channels.
This is the primary source of infidelity in the multichannel architecture, which we now analyze.
\red{Here, we do not explicitly include potential disorder caused by lever-arm fluctuations, trapped charge, or alloy disorder between the channels. 
For our initial simulations, these effects may simply be absorbed into the definitions of the detuning parameter  $\varepsilon$ and the tunnel coupling $t_c$.
In other simulations, described below, we explicitly include fluctuations of the detuning parameter. 
We also consider tunnel coupling fluctuations, finding that they have a weaker effect than detuning fluctuations, except in extreme cases. 
From an experimental perspective, we can view these locally varying parameters as being characterizable. 
For example, one could use the shuttler itself as a probe, as demonstrated in~\cite{Volmer:2024p61}.
Following characterization, the pulse sequence used to perform a transfer operation should be modified at each transfer location, to account for the observed fluctuations.}

We first investigate multichannel shuttling in a ``paused'' protocol, in which the longitudinal conveyor motion is temporarily halted to allow for tunneling between the channels.
To evaluate the success of channel transfer, we perform simulations of Eq.~(\ref{eq:total_ham}) as a function of time $\tau$, while modulating $\varepsilon$ and $t_c$ according to the following schedules:
\begin{gather} 
\label{eq:single_channel_params}
\varepsilon(\tau) = \varepsilon_0 \left( -1 + 2 \tau / \tau_{\text{tot}} \right) , \\
\label{eq:single_channel_params2}
t_c (\tau)  = t_0 \sin \left(\pi \tau /  \tau_\text{tot} \right) ,
\end{gather}
as illustrated in Fig.~\ref{fig:2}(c).
In the simulations, we assume the dot is initialized into its ground state in the left channel, $|\psi (\tau=0) \rangle = |L,g \rangle$.
The detuning parameter then transitions from $-\varepsilon_0$ to $\varepsilon_0$ over the transfer time $\tau_\text{tot}$.
During this same period, the tunnel coupling is modulated smoothly from zero, to its maximum value $t_0$, and back to zero.
\red{(For the ``paused'' shuttling protocol, $t_c$ modulation is not needed; however, for the ``moving'' protocol considered below, it is important to turn off the tunnel coupling, to reduce transfer errors -- particularly errors arising from lever-arm fluctuations, described in Appendix~\ref{app:landscapes}.)} 
\red{We note that optimal control techniques~\cite{Ban:2012p206602,Khomitsky:2012p125312,Ban:2014p6258} can help to achieve better transfer fidelities, as compared to the schedules shown in Eqs.~(\ref{eq:single_channel_params}) and (\ref{eq:single_channel_params2}); however, we do not explore this possibility here.}

Below, we define the transfer fidelity as the fraction of the wavefunction remaining in the ground valley state in the right channel at the end of the procedure: $F = |\langle R,g | \psi(\tau_\text{tot} + 5 \text{ ns}) \rangle |^2$, where we wait an additional \SI{5}{\nano\second} to allow the simulations to stabilize.
\red{Defined in this way, any deviations from $F=1$ represent leakage errors, since the logical states are encoded entirely within the spin space of $|R,g\rangle$.
Although we do not distinguish between different leakage channels here, we note they are dominated by the occupation of $|R,e\rangle$.
Such excitations could potentially be correctable, for example, by encouraging fast valley relaxation~\cite{Losert:2024p040322}, although we do not explore this possibility here.
Target values for $F$ depend on considerations such as error-correction codes~\cite{Kunne:2024p4977}, the size of the shuttler, and the density of locations with low valley splittings, which we do not try to estimate here.
In our numerical analysis below, we simply define channel-transfer ``success'' as $1-F\leq 10^{-3}$, representing an arbitrary but low infidelity value.}

To begin, we perform simulations of channel transfer with no knowledge of the valley-state landscape.
Since we are working in the ADD regime, the valley phases are randomized at any given location.
As noted above, the general case of $\delta\phi\neq 0$ can cause valley excitations; in particular, $\delta\phi=\pm\pi$ always causes excitations.
Away from this worst-case scenario, excitations \red{are enhanced when either channel has a low valley splitting, but} can be suppressed by adiabatic operation, as determined by the values of $\varepsilon_0$, $t_0$, $\tau_\text{tot}$, \red{$|\Delta_L|$, and $|\Delta_R|$}.
In Fig.~\ref{fig:2}(d), we plot the transfer success probability $P_\text{suc}$ as a function of $\varepsilon_0$ and $t_0$, for a typical fixed transfer time of $\tau_\text{tot} = 10$~ns.
\red{As noted above,} we define ``success'' as a transfer infidelity of $ 1 - F \leq 10^{-3}$.
Each data point is obtained by averaging the results from 10,000 simulations.
Here and throughout this work, the randomized valley couplings $\Delta_L$ and $\Delta_R$ are chosen from a complex normal distribution centered around zero, \red{giving valley splittings $E_v=2|\Delta|$ that follow a Rayleigh distribution~\cite{Losert:2023p125405} and valley phases from a uniform distribution on $[0,2\pi)$. 
We choose} a standard deviation of $\sigma_\Delta = $ \SI{56.4}{\micro\electronvolt} and an average valley splitting of \SI{100}{\micro\electronvolt}, typical of recent experiments \cite{Wuetz:2022p7730, Esposti:2024p32}.
These settings can be achieved, for example, by adding a small amount of Ge to the quantum well.
In the figure, we observe low success probabilities in the limit of low $t_0$, since the dot does not have sufficient time to tunnel between channels.
We also observe poor success in the limit of low $\varepsilon_0$, since the final state remains hybridized between the channels.
Higher success rates are obtained more generally for large $t_0$ and $\varepsilon_0$ values, although very large $\varepsilon_0$ values can cause a slight suppression of $P_\text{suc}$ when the Landau-Zener velocity is very high. 
Nonetheless, for $t_0 \gtrsim 50$~\SI{}{\micro\electronvolt} and $\varepsilon_0 \gtrsim 150$~\SI{}{\micro\electronvolt}, we generally observe success rates $\gtrsim 85$~\%.
In a second set of simulations, we therefore choose parameters within this range, with $t_0\in(50,150)$~\SI{}{\micro\electronvolt} and $\varepsilon_0\in(500,1000)$~\SI{}{\micro\electronvolt}.
As shown in Fig.~\ref{fig:2}(e), longer transfer periods yield more-adiabatic behavior and higher success rates for all shuttling parameters, as expected, with success rates $\gtrsim 95$~\% in many cases.
We also note that if the valley landscape can be mapped out before performing a channel transfer, the location of the transfer can be adjusted to improve the transfer success probability.

We have shown, above, that the shuttling parameters $\varepsilon_0 \sim 500$~\SI{}{\micro\electronvolt} and $t_0 \geq 100$~\SI{}{\micro\electronvolt} provide a good working point for high-fidelity channel transfer.
We now perform electrostatic simulations of the device shown in Figs.~\ref{fig:1}(b) and \ref{fig:2}(a), to confirm that the parameters assumed in Eqs.~(\ref{eq:single_channel_params}) and (\ref{eq:single_channel_params2}) are physically reasonable.
(Note that the clavier gates C play the role of plunger gates here, while the screening gates S1-S3 act as barrier gates.)
We first choose a set of clavier gate voltages, using the sinusoidally varying scheme common to conveyor-mode shuttling experiments~\cite{Langrock:2023p020305, Seidler:2022p100, Xue:2024p2296, Kunne:2024p4977}:
\begin{equation}
V_{i}(\tau) = V_\text{amp} \cos \left( \Omega_x \tau + \delta \theta^x_{i} \right) . \label{eq:Vi1D}
\end{equation}
Here, $V_i$ is the voltage applied to gate $i$ and we set $\Omega_x=0$ for a paused-style channel transfer.
Choosing phase shifts of $\delta\theta_{i}^x=\pi/2$ on successive clavier gates yields a ``unit cell'' of four gates ($V_i=V_{i+4}$), as illustrated in Fig.~\ref{fig:1}(b).
For definiteness, we choose a physically reasonable voltage amplitude of $V_\text{amp}=100$~mV in our simulations and an overall phase that centers the dot halfway between two clavier gates.
Electrostatic simulations are performed using the MaSQE software package~\cite{Anderson:2022p065123}. 
Here, we vary the voltages $V_\text{S1}$-$V_\text{S3}$ on the screening gates S$_1$-S$_3$, which differs from conventional double dots, where tuning is normally performed with plunger gates. 
We compute $\varepsilon$ and $t_c$, as described in Appendix~\ref{app:electrostatics}, searching for $\varepsilon$ and $t$ parameter values consistent with Fig.~\ref{fig:2}(d).
We find that the protocols given by Eqs.~(\ref{eq:single_channel_params}) and (\ref{eq:single_channel_params2}) may be implemented with voltage changes no larger than $\Delta V=100$~mV on any of the screening gates.
Two resulting 2D potential-energy profiles are shown in Fig.~\ref{fig:2}(f), for the cases of $\tau=0$ (top) and $\tau=\tau_\text{tot}/2$ (bottom).
Vertical linecuts through the data, along the line $x=0$, are plotted in Fig.~\ref{fig:2}(g), indicating regions with low $t_c$ (blue) and high $t_c$ (red).
A blown-up region at the bottom of the red curve is shown in the inset.
Note that it is not possible to fully extinguish $t_c=0$ at the endpoints of the transfer protocol, due to the nonzero overlap of wave functions in the two channels. 
However, we obtain $t_c =10^{-6}$~\SI{}{\micro\electronvolt} here, which does not significantly degrade the transfer fidelity.

\red{It is important to compare the effects of disorder-induced fluctuations to these electrostatically defined potential variations.
We first note that the electrostatic variations are of order 10-20~meV, with corresponding orbital confinement energies along $\hat {\bm x}$ on the order of $\sim$2~meV.
The fluctuating energy landscape due to random-alloy disorder is much smaller, with variations $\lesssim 0.1$~meV.
For the simulations described above, this justifies our approach of absorbing the fluctuations into the Hamiltonian parameters, assuming they can be characterized at each transfer location. 
This is more challenging in the simulations described below, where we explicitly include the fluctuations in the Hamiltonian. 
Since disorder-induced potential variations are still small in this case, compared to electrostatic effects, we treat these fluctuations perturbatively, as described in Appendix~\ref{app:landscapes}.}

\begin{figure*}[t] 
	\includegraphics[width=14cm]{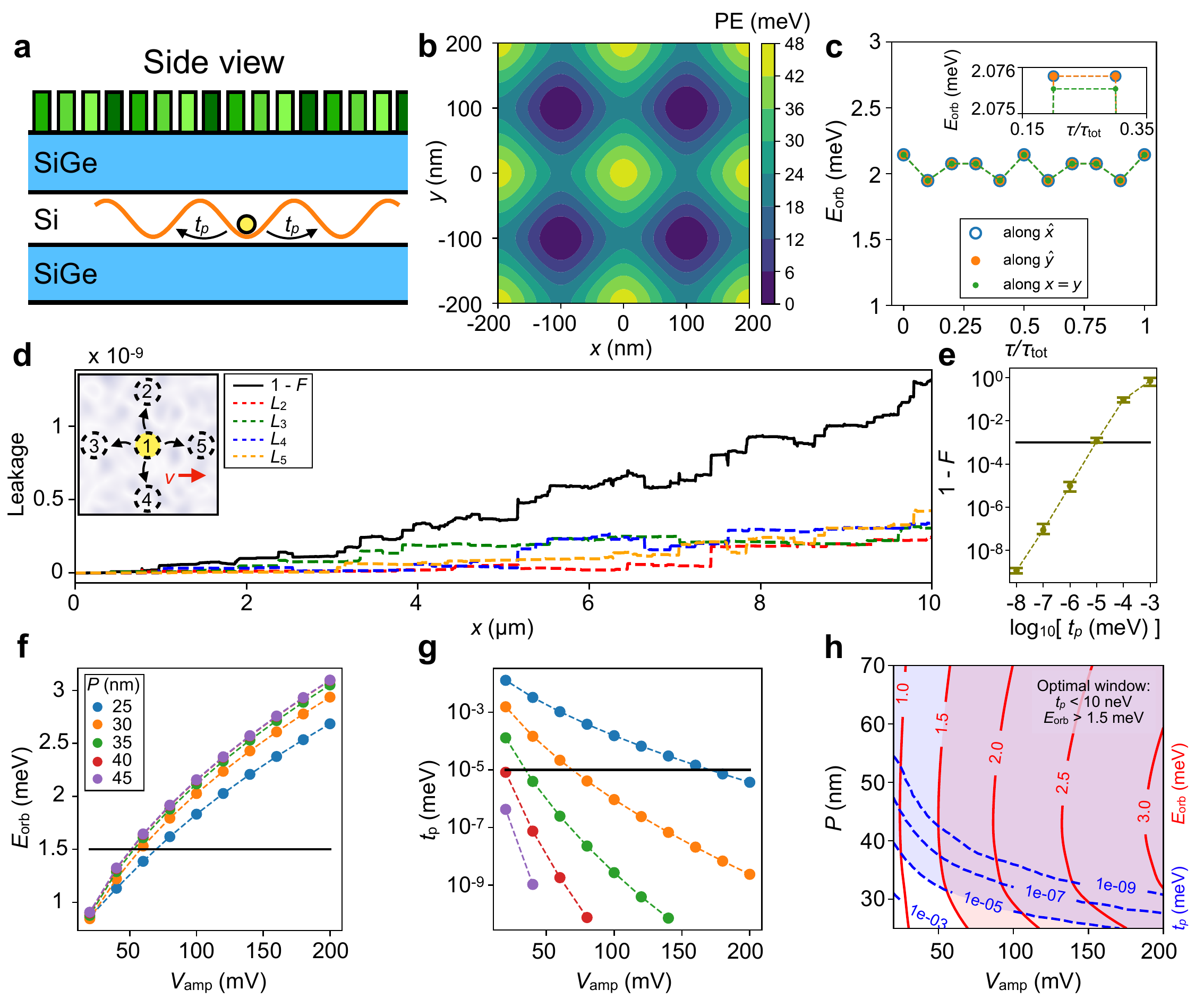}
	\centering
	\caption{
    2D shuttling simulations.
(a) Schematic side view of a shuttling device, showing a Si/SiGe quantum well with top-gate electrodes.
For a 2D shuttler, the top (``clavette’’) gates are formed into 2D unit cells, indicated by shading.
By applying sinusoidally varying voltages to these gates, we obtain a 2D array of moving potential pockets in the quantum well (orange curve), capable of moving an electron in any direction.
The couplings $t_p$ induce tunneling events between neighboring pockets.
(b) Electrostatic simulations of the potential energy (PE) in the quantum well. (See Appendix~\ref{app:electrostatics}.)
(c) Orbital confinement energies of the moving potential pockets for motion along directions defined by $x=0$, $y=0$, or $x=y$, showing stable, omnidirectional transport.
(Inset shows a blown-up region.)
(d) The leakage probability $1-F$ to neighboring pockets, Eq.~(\ref{eq:leakage}), is determined as a function of shuttling distance $x$, for the case of small tunnel couplings $t_p = 10^{-8}~\SI{}{\milli\electronvolt}$, which are typical for this system.
Simulations include the effects of potential and valley-splitting fluctuations (shown in the inset as shading variations for a typical, randomized landscape), and charge-state collapse (see main text).
The total leakage out of the central pocket is plotted as a solid line, while leakage into individual pockets (see inset) is shown as dashed lines.
Note the low leakage scale of $10^{-9}$ found here, indicating that pocket leakage should not be a problem under normal operating conditions.
(e) The total leakage $1-F$ at a shuttling distance of $x=10~\SI{}{\micro\meter}$ is plotted as a function of $t_p$, based on simulations averaged over five disorder realizations. 
Here, we assume a dot radius of $l_\text{dot} = 14~\SI{}{\nano\meter}$, corresponding to a typical orbital splitting of $E_\text{orb} = 2~\SI{}{\milli\electronvolt}$. 
For the simulations in (b)-(e), we set $V_\text{amp} = 100~\SI{}{\milli\volt}$, $P = 50~\SI{}{\nano\meter}$, and $W = 5~\SI{}{\nano\meter}$. 
(f) Orbital excitation energies $E_\text{orb}$, as a function of sinusoidal voltage amplitude $V_\text{amp}$ [defined in Eq.~(\ref{eq:Vi2D})], for the indicated gate pitches $P$.
A reasonable threshold of $E_\text{orb}=1.5$~meV~\cite{Langrock:2023p020305} is indicated (black line), above which orbital excitations are strongly suppressed.
(g) The tunnel coupling $t_p$ between neighboring potential pockets, as a function of $V_\text{amp}$, for several different $P$ values, using the same color scheme as (f).
A reasonable threshold of $t_p=10^{-5}$~meV is indicated (black line), below which we can assume tunneling to nearby pockets is strongly suppressed.
(h) The same results as (f) and (g), combined into a contour plot.
The optimal operating window is shaded purple, indicating that orbital and pocket leakage can be strongly suppressed over a wide range of parameters.} 
	\label{fig:3}
\end{figure*}

We also consider a second, ``moving'' protocol in which the qubit is not paused while performing a channel transfer.
In this case, we assume a constant longitudinal velocity of $v_x = 2\Omega_xP/\pi$, where $P$ is the clavier gate pitch defined in Fig.~\ref{fig:1}(b), and $\Omega_x > 0$.
We again simulate the time-evolution of Eq.~(\ref{eq:total_ham}).
However, since the shuttling electron moves across a nonuniform valley terrain, the basis transformation $U_v$ is no longer static, which introduces a dynamical correction to the time evolution: $\tilde H_\text{eff} = \tilde H - i \hbar U_v \dot U_v^\dag$. 
Similarly, 
we cannot ignore the spatially varying potential disorder that was previously absorbed into $\varepsilon$.
The Hamiltonian therefore acquires corrections of the form $\mathcal{P}_L \delta \varepsilon_L(\tau) + \mathcal{P}_R \delta \varepsilon_R(\tau)$, where $\delta \varepsilon_{L(R)}$ describe potential-energy fluctuations in the left (right) channels.
We consider fluctuations arising from two sources: $\delta \varepsilon_{L(R)} = \delta \varepsilon_\text{alloy}^{L(R)} + \delta \varepsilon_\text{gate}^{L(R)}$.
The potential fluctuations due to alloy disorder ($\delta \varepsilon_\text{alloy}$) are assumed to be normally distributed, to be uncorrelated between channels (i.e., $\delta \varepsilon_\text{alloy}^L \neq \delta \varepsilon_\text{alloy}^R$), and to have a characteristic size given by $\sigma_\Delta$.
Potential fluctuations from all other sources are absorbed into $\delta \varepsilon_\text{gate}$ and modeled as normally distributed random fields with a characteristic magnitude of \SI{1}{\milli\electronvolt} (see Appendix~\ref{app:landscapes}) and a correlation length along the channel given by the gate pitch, taken to be $P = 70$~\SI{}{\nano\meter}.
We then explore two regimes for $\delta \varepsilon_\text{gate}$: (1) a ``correlated'' regime, where $\delta \varepsilon_\text{gate}^L = \delta \varepsilon_\text{gate}^R$ (e.g., in the case of lever-arm fluctuations, due to imperfections in gate size), and (2) an ``uncorrelated'' regime, where $\delta \varepsilon_\text{gate}^L \neq \delta \varepsilon_\text{gate}^R$.
In reality, both types of behavior are likely to be present.
\red{We have also considered the possibility of tunnel coupling disorder, making use of the log-normal disorder model described in Appendix~\ref{app:landscapes}.
As a test, we consider fluctuation distributions defined by the standard deviations $\sigma_t/t_0=5\%$, 30\% and 200\%, combined with detuning disorder, as described above. 
Taking 200 different disorder realizations for each of these cases, we find that detuning disorder dominates the results, except in the extreme case of $\sigma_t/t_0=200\%$, where the computed success probabilities are reduced by a small fraction. 
In these simulations, we therefore focus only on detuning fluctuations.}

We first consider the correlated regime, taking $\varepsilon_0 = 500$~\SI{}{\micro\electronvolt} and $t_0 = 100$~\SI{}{\micro\electronvolt}, and choosing typical values of $\tau_\text{tot}$ between 10 and \SI{50}{\nano\second}.
In Fig.~\ref{fig:2}(h), we plot success probabilities as a function of $\tau_\text{tot}$ for realistic shuttling velocities $v_x \in \{1, 5, 10\}$~\SI{}{\meter\per\second}, averaged over 200 disorder landscapes.
We find the success probability to be significantly reduced at higher velocities, as the moving dot is more likely to encounter regions where $\delta \phi \approx \pi$, resulting in valley excitations.
Nonetheless, for slower shuttling velocities, $\sim 1$~\SI{}{\meter\per\second}, we can achieve relatively high success probabilities of $\sim 90$~\%.

Next, we consider the uncorrelated regime. 
Setting $v_x = 1$~\SI{}{\meter\per\second}, we again perform simulations with $\tau_\text{tot}$ between 10 and \SI{50}{\nano\second}.
In Fig.~\ref{fig:2}(i), when $\varepsilon_0 = 500$~\SI{}{\micro\electronvolt}, we observe very poor success probabilities, since the detuning pulse is dominated by potential disorder between the channels.
Increasing $\varepsilon_0$ to \SI{5000}{\micro\electronvolt} slightly improves the situation, although the results obtained for $P_\text{suc}$ are still lower than those observed in the correlated regime.
Overall, these results highlight the additional complications that the ``moving'' protocol poses for channel transfer, since it would require careful calibration of microscopic disorder and gate pulses to achieve high-fidelity operation.

In summary, our results suggest that channel transfer is more difficult for moving dots than for paused shuttlers.
However, paused schemes are not highly scalable, since all the electrons in a shuttling channel must be paused simultaneously, running the risk of accumulated dephasing errors.
The electrons must also be transferred between channels simultaneously, unless the shuttler can be broken up into smaller segments, adding complexity to the device.
Together, these drawbacks motivate the development of a fully 2D conveyor-mode shuttling approach that avoids tunneling-based processes altogether.

\section{Fully 2D shuttling}
We now consider the 2D conveyor-mode shuttling scheme, illustrated schematically in Figs.~\ref{fig:1}(c) and \ref{fig:3}(a).
This is a natural extension of the 1D shuttler, which now comprises a 2D unit cell of ``clavette'' gates, tiled to cover the heterostructure.
Here, we consider a $4 \times 4$ unit cell with 16 independent signal lines.
Such devices cannot be easily fabricated using overlapping gates, and will likely require industrial fabrication techniques, such as etched deposition and vertical vias~\cite{Ha:2022p1443, Acuna:2024p044057, George:2025p793}.

To achieve conveyor-mode control in 2D, we apply the sinusoidally varying gate voltages,
\begin{equation}
V_{ij}(\tau) = \frac{V_\text{amp}}{2} \left[ \cos \left( \Omega_x \tau + \delta \theta^x_{ij} \right) +  \cos \left(\Omega_y \tau + \delta \theta^y_{ij} \right) \right] , \label{eq:Vi2D}
\end{equation}
where the voltage $V_{ij}$ is applied to the clavette gate indexed by $(i,j)$.
Parameters $\Omega_{x(y)}$ and $\delta \theta_{ij}^{x(y)}$ are defined analogously to Eq.~(\ref{eq:Vi1D}), with phase shifts of $\pi/2$ applied between nearest-neighbor gates along the $x$ and $y$ axes.
Omnidirectional control of the shuttler is achieved by independently tuning the parameters $\Omega_{x(y)}$, with corresponding velocities $v_{x(y)} = 2\Omega_{x(y)}P/\pi$.
The shuttling direction (i.e., angle), measured from the $x$-axis, is then given by $\varphi_\text{sh} = \tan^{-1} ( \Omega_y / \Omega_x)$.
\red{For constant values of $v_{x(y)}$, the resulting trajectories are linear.
A more interesting example is given by a trajectory that detours around a region of low valley splitting.
We consider a semicircular path of radius $R$ that traverses around the defect clockwise at a constant velocity $v=\pi R/\tau_\text{circ}$.
For the detour period $\tau\in (0,\tau_\text{circ})$, we adopt the pulse scheme
\begin{equation}
v_x=v\sin(\pi\tau/\tau_\text{circ}), \quad
v_y=v\cos(\pi\tau/\tau_\text{circ}) ,
\end{equation}
where the corresponding voltages are given by Eq.~(\ref{eq:Vi2D}), with the definition $\Omega_{x(y)}=\pi v_{x(y)}/2P$.
Equation~(\ref{eq:Vi2D}) then describes 16 different voltage expressions, which differ only in their phases $\delta\theta_{i,j}^{x(y)}$.
As noted in \cite{Losert:2024p040322}, a detour radius of $R=50$~nm should be sufficient for typical dot sizes.}

We first simulate the 2D shuttling geometry electrostatically, as described in Appendix~\ref{app:electrostatics}, assuming an amplitude of $V_\text{amp} = 100$~mV, a gate pitch of $P = 50$~nm, and an intergate spacing of $W = 5$~nm, obtaining the 2D potential pockets shown in Fig.~\ref{fig:3}(b).
These potentials are taken as inputs in a 2D Schr\"odinger solver, to determine the 2D ground and excited pocket (orbital) wavefunctions and their energy values, which are used in the calculations described below. 
The pocket confinement potentials are also used to compute the interpocket tunnel couplings $t_p$, as described in Appendix~\ref{app:electrostatics}.
In Fig.~\ref{fig:3}(c), we plot the orbital excitation energies for an electron confined to a single potential pocket, as a function of time, for three linear trajectories: along $\hat x$ (setting $\Omega_y = 0$), along $\hat y$ (setting $\Omega_x = 0$), and along $y = x$ (setting $\Omega_x = \Omega_y$).
For the same values of $V_\text{amp}$, $P$, and $W$ used in Fig.~\ref{fig:3}(b), we obtain orbital excitation energies above 1~\SI{}{\milli\electronvolt} over the whole oscillation period, regardless of shuttling direction, which is important for suppressing orbital excitations.

If the valley-state landscape is well-characterized and the shuttling path is chosen to avoid valley-splitting minima, valley excitations can largely be avoided.
\red{Such a strategy was explored in detail in~\cite{Losert:2024p040322}, where detour paths as large as 100~nm were considered.
Those results apply directly to the present work, so we briefly summarize them here.
In \cite{Losert:2024p040322}, the focus was on the key problem of valley excitations caused by disordered energy landscapes arising from alloy disorder, although other effects like charge noise and magnetic noise were also considered.
Strategies for addressing this problem were studied, including adding Ge to the quantum well, and varying the vertical electric field, the shuttling velocity, the
shape and size of the dot, and the shuttling path.
It was shown that \emph{combinations of strategies} can provide excellent shuttling fidelities $>$99.99\%, over long 10~$\mu$m trajectories. 
Additionally, it was shown that tradeoffs involving the shuttling speed lead to optimal shuttling velocities of order of several~m/s.}

\red{We do not repeat the calculations of \cite{Losert:2024p040322} here.}
Instead, we focus on whether some new type of behavior -- directly related to 2D shuttling  -- emerges as a new, prevalent leakage mechanism.
Specifically, we consider the consequences of unwanted tunneling to neighboring pockets in the 2D shuttling array, mediated by the tunnel couplings $t_p$, as indicated in Fig.~\ref{fig:3}(a).
In principle, $t_p$ can be suppressed by increasing the gate amplitude $V_\text{amp}$, which determines the barrier height between the pockets; however, large ac potentials produce excessive Ohmic heating, especially in large shuttlers, and should be avoided.
Alternatively, $t_p$ may be suppressed by increasing the gate pitch $P$, which moves the pockets farther apart; however, \red{a very large gate pitch can lower the dot confinement, as observed in Fig.~\ref{fig:3}(h), increasing the chance of orbital excitations in some cases.}

To explore these tradeoffs, we first perform time-evolution simulations.
We consider a basis set of five localized potential pockets: a central pocket and its four nearest neighbors, as illustrated in the inset of Fig.~\ref{fig:3}(d).
The electron envelope functions are assumed to have a gaussian waveform, due to their approximately parabolic confinement; this determines their overlap with the random valley-coupling landscape (see below and in Appendix~\ref{app:landscapes}) and their electron-phonon matrix elements (see Appendix~\ref{app:relaxation}).
Within each pocket labeled $|d_j\rangle$, where $j=1$ to 5  and $j=1$ refers to the central pocket, we include two valley states labeled $|z_\pm\rangle$, resulting in a 10-level system: $\{ |z_\pm,d_j\rangle\}$. 
The Hamiltonian of the five-pocket system at a fixed time-step is given by $H = H_\text{os} + H_\text{hop} + H_\text{val}$, consisting of an on-site energy term, 
\begin{equation}
    H_\text{os} = \sum_{s=\pm} \sum_{j=1}^5 \varepsilon_j | z_s, d_j \rangle \langle z_s, d_j | ,
    \label{eq:onsite}
\end{equation} a hopping term responsible for coherent tunneling,
\begin{equation}
    H_\text{hop} = \sum_{s=\pm} \sum_{j=2}^5 t_p^{j} | z_s, d_1 \rangle \langle z_s, d_j | + \mathrm{h.c.}\,,
    \label{eq:hopping}
\end{equation} and a term that couples the $z_\pm$ valley states,
\begin{equation}
    H_\text{val} = \sum_{j=1}^5 \Delta_j |z_+, d_j \rangle \langle z_-, d_j | + \text{h.c.} 
    \label{eq:valcoup}
\end{equation} 
Similar to the two-channel simulations, we consider disordered potential landscapes, resulting in location-dependent values of $\varepsilon_j$, $t_p^j$, and $\Delta_j$, which we generate randomly from statistical distributions, as described in Appendix~\ref{app:landscapes}.
As the five-pocket system traverses the shuttling device, it samples these landscape variations.
Below, we consider the instantaneous eigenstates of $H$, which form hybridized, delocalized states, denoted by $|m\rangle$ and $|n\rangle$.

Tunneling between neighboring dots involves both coherent and decoherent processes.
We therefore consider Lindblad collapse operators $L_{nm} = |m\rangle\langle n|$, describing transitions between the $n$th and $m$th instantaneous eigenstates. 
Because of the pocket motion described by Eq.~(\ref{eq:Vi2D}), the Hamiltonian parameters $\varepsilon_j$, $t_p^j$, and $\Delta_j$, the instantaneous eigenstates $|m\rangle$ and $|n\rangle$, and the collapse operator $L_{nm}$ all depend on the shuttling time $\tau$, which we suppress here, for brevity.
We initialize the simulation into the ground valley state of the central pocket, $|g\rangle$, such that $\rho(\tau = 0) = |g, d_1\rangle\langle g, d_1 |$.
Here, $ |g,d_1\rangle$ and $ |e,d_1\rangle$ are defined as the eigenstates of $H_\text{val}$ for the central pocket.
We then perform time-evolution simulations of the Lindblad master equation
\begin{equation}
\begin{aligned}
\dot \rho = &- \frac{i}{\hbar} [H, \rho] \\ &+ \underset{n \neq m}{\sum_{n,m=1}^{10}} \Gamma_{nm} \left( L_{nm} \rho L_{nm}^\dag - \frac{1}{2} \{ L_{nm}^\dag L_{nm}, \rho \}\right),
\end{aligned}
\label{eq:Lindblad}
\end{equation} 
where the relaxation rates $\Gamma_{nm}$ are set by electron-phonon interactions, which dominate the decay processes for large energy splittings. (See Appendix~\ref{app:time-evolution}.)
We define the tunneling fidelity of a simulation as the probability that the electron remains in the central dot after a shuttling period of $\tau_\text{tot}$ and a distance of 10~\SI{}{\micro\meter}: 
\begin{equation}
    F = \langle g, d_1| \rho(\tau_\text{tot}) | g, d_1 \rangle + \langle e, d_1| \rho(\tau_\text{tot}) | e, d_1 \rangle ,
    \label{eq:leakage}
\end{equation}
and we define $1-F$ as the leakage out of the central dot.
As intended, this definition accounts for 2D tunneling effects, but not valley excitations.

Typical simulation results are presented in Fig.~\ref{fig:3}(d), where the occupations of the individual pockets and the total pocket leakage are shown as a function of the shuttling distance. 
The results show abrupt jumps separated by regions of slowly changing leakage. 
Both of these behaviors are associated with tunneling: the former is caused by the hybridization of localized states in neighboring pockets when their energy levels cross, while the latter describes incoherent, phonon-mediated relaxation into a nearby pocket, with a rate proportional to the phonon density of states, which increases with the level splitting. (See Appendix~\ref{app:time-evolution}.)
In Fig.~\ref{fig:3}(e), we plot the average shuttling infidelities as a function of the tunnel coupling $t_p$; here, each point is averaged over five different simulations. 
Importantly, we note that the average infidelity falls below $10^{-3}$ for tunnel couplings of $t_p<10^{-5}$~\SI{}{\milli\electronvolt}.

\begin{figure*}[t] 
	\includegraphics[width=12cm]{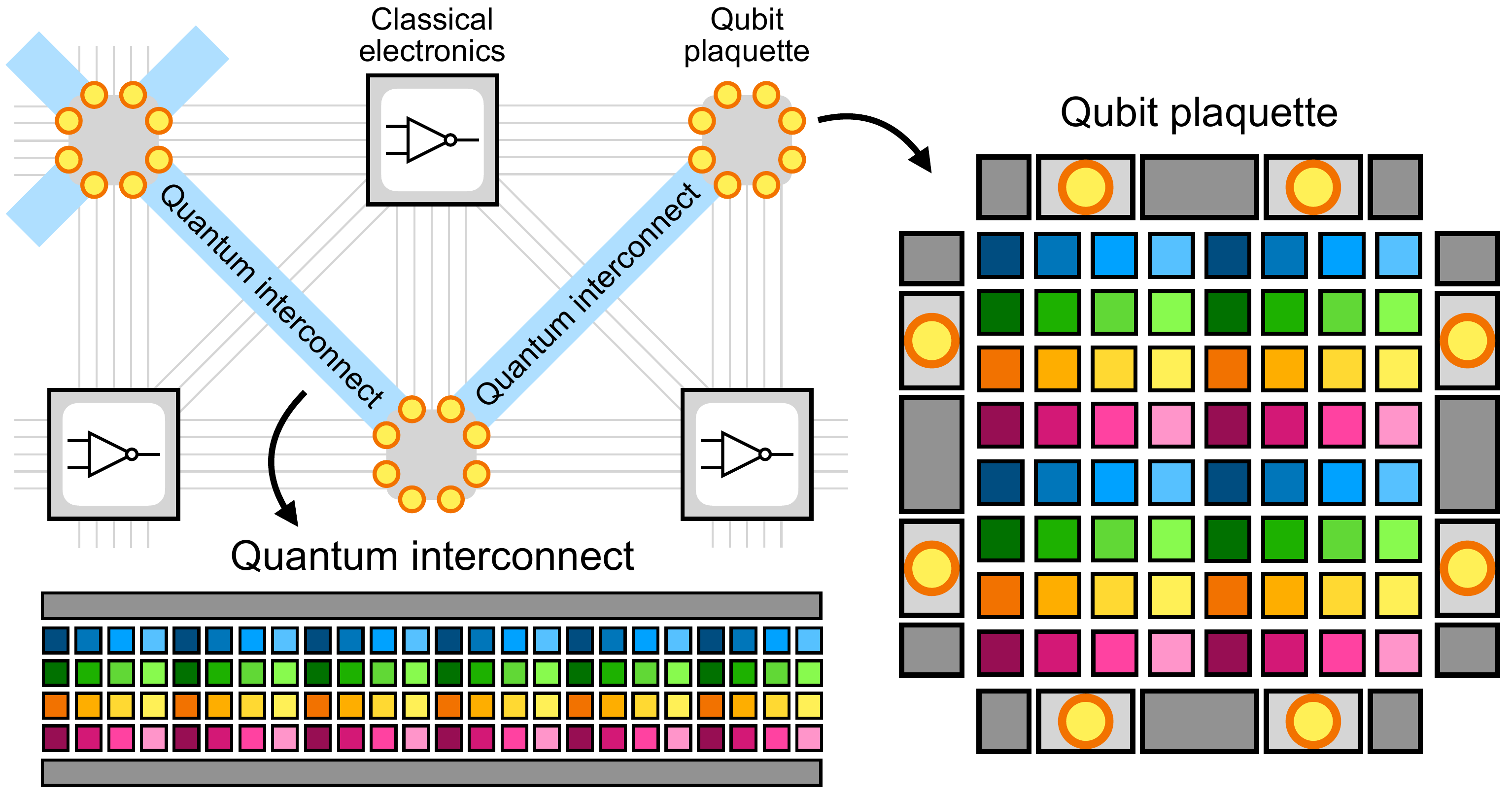}
	\centering
	\caption{
    A proposed quantum computing architecture based on 2D shuttling.
The top-left shows a schematic illustration of a \red{modular architecture}, with similarities to~\cite{Vandersypen:2017p34}, incorporating three distinct technologies: qubit plaquettes (comprised of qubits, readout and control electronics, arranged around the periphery of a 2D shuttler), quantum interconnects (also comprised of 2D shuttlers), and classical control electronics interspersed between the qubit plaquettes.
All-to-all connectivity is enabled within a single plaquette, while  the quantum interconnects allow electrons to shuttle around regions of low valley splitting.
Although the illustration here is schematic, we envision future architectures with higher numbers of qubits, specialized structures for implementing one or two-qubit gate operations, and seamless connections between plaquette shuttlers and interconnects.
    }
	\label{fig:4}
\end{figure*}

Thus, it is important to lower $t_p$ 
through appropriate choices of $V_\text{amp}$ and $P$.
However, these choices can also lower $E_\text{orb}$ (as noted above), potentially causing leakage due to orbital excitations.
To investigate this problem, in Figs.~\ref{fig:3}(f) and ~\ref{fig:3}(g), we plot $E_\text{orb}$ and $t_p$, respectively, as functions of $V_\text{amp}$, for five different $P$ values.
(These quantities are determined via electrostatic simulations, as described in Appendix~\ref{app:electrostatics}.)
We combine this information in Fig.~\ref{fig:3}(h) by plotting $E_\text{orb}$ and $t_p$ contours as a function of $P$ and $V_\text{amp}$.
Here, we indicate with purple shading the high-fidelity operating regime where $t_p < 10^{-5}$~\SI{}{\milli\electronvolt} \emph{and} $E_\text{orb} > 1.5$~\SI{}{\milli\electronvolt}, where the latter was suggested in~\cite{Langrock:2023p020305}. 
In this way, we identify the high-fidelity shuttling regime as having parameters $P\gtrsim 35$~nm and $V_\text{amp}\gtrsim 75$~mV. 
\red{The observation of low leakage errors in this system is not surprising, based on expectations from 1D shuttlers~\cite{Langrock:2023p020305}; however, the robustness of shuttling fidelities in Fig.~\ref{fig:3}(h) over a wide range of 2D system parameters is reassuring.}

In summary, we have shown that our proposed 2D shuttler enables rapid, omnidirectional transport.
The 2D valley-splitting landscape can then be mapped out, similar to~\cite{Volmer:2024p61}, and a path can be chosen to avoid valley excitations, while reasonable values of $P$ and $V_\text{amp}$ can be adopted to suppress inter-pocket tunneling.
For a unit cell of $4\times 4$ clavette gates, the scheme requires only 16 independent control lines, which does not present an extravagant wiring cost.
When many unit cells are connected together, this enables transverse shuttling shifts of $\Delta y$$>$\SI{100}{\nano\meter} (or more), as required for high-fidelity operation~\cite{Losert:2024p040322}.
The full wiring cost depends on the required versatility of the shuttler.
For example, if simultaneous bidirectional transport is needed for different electrons, then the gate array needs to be segmented, with a corresponding increase in wiring.

\section{Discussion and Conclusions}
We have proposed two schemes to enhance the transverse maneuverability of an electron shuttler in a Si/SiGe quantum well, to address the problem of very low valley splittings, which are likely to be encountered along a long shuttling trajectory.
To achieve high-fidelity operation, shifts of $\sim$100~nm are required, which is not possible in existing shuttlers.
Both of our schemes require, as a first step, to enhance the average valley splitting via heterostructure engineering, e.g., by adding a small amount of Ge to the quantum well. 
The first scheme extends the conventional single-channel shuttler to two channels (or more), separated by a tunable tunnel barrier, enabling the desired 100~nm channel shift.
We have simulated two control protocols for such multichannel devices: 
(1) A ``paused’’ scheme, in which tunneling occurs while the shuttling is halted. 
In this case, the fidelity is limited by valley-state excitations as the electron tunnels between shuttling channels.
We consider this mode to be poorly scalable because the pause applies to all electrons present in the shuttler.
(2) A ``moving’’ scheme, in which tunneling occurs while shuttling is in progress.
While this mode does not suffer from the same scaling challenges, the fidelity is more severely limited by valley excitations, due to the dynamic disorder encountered while shuttling.
Overall, multichannel shuttling is found to be a promising approach for near-term experiments, since it can be implemented using standard fabrication techniques; however, the resulting fidelities may not be appropriate for large-scale quantum-computing applications.
A more scalable approach is given by the 2D shuttler, based on a periodic tiling of 2D unit cells containing clavette gates.
In this scheme, the greatest threat to fidelity also arises from valley-state excitations, even when shuttling paths are chosen to avoid valley minima.
However, previous work suggests that the resulting fidelities can be high under these conditions~\cite{Losert:2024p040322}.

Based on these encouraging results, we envision a further extension of the 2D shuttler to a full shuttling-based quantum computing architecture, as illustrated in Fig.~\ref{fig:4}.
Such plaquette geometries address multiple challenges faced by quantum-dot quantum computers~\cite{Vandersypen:2017p34}, including wiring fanout and classical, on-chip control. 
The main problem addressed so far in this paper pertains to the fidelity of the quantum links between plaquettes, as illustrated at the bottom of the figure; here, the 2D shuttler provides a means of avoiding valley excitations.
The challenge addressed on the right-hand side of the figure relates to qubit \emph{connectivity}. 
It is well known that the natural two-qubit gates between spin qubits, based on the exchange interaction, are extremely short-ranged ($\sim$10~nm)~\cite{Burkard:2023p025003}.
As such, these interactions allow only nearest-neighbor qubit gate operations, which are known to have poor scaling properties, particularly with regards to quantum error correction~\cite{Bravyi:2024p778}.
However, even small improvements in connectivity can provide significant improvements in quantum error correction~\cite{Bravyi:2024p778}.
In Fig.~\ref{fig:4}, we imagine using a 2D shuttler as a mediator for all-to-all connections between qubits in a plaquette.
In this scheme, localized qubits are envisioned around the periphery of the shuttler, although they could also be placed in the interior. 
Two-qubit gates may now be implemented by transferring a qubit onto the shuttler and physically transporting it to a target qubit, where a gate operation takes place.
When the operation is complete, the qubit may be transported to its original site, or elsewhere.
Since the shuttler can transport many electrons simultaneously, the issue of scalability in this architecture is reduced to a scheduling problem.
A shuttling-based architecture therefore provides an interesting and scalable alternative to conventional, low-connectivity architectures.

\section*{Acknowledgments}
We are grateful to Jan Krzywda, Michael Wolfe, Talise Oh, Emily Eagen, and Ben Woods for helpful discussions.
R.\ N\'{e}meth's work on this project was supported by a Fulbright Program grant sponsored by the Bureau of Educational and Cultural Affairs of the United States Department of State and administered by the Institute of International Education and the Hungarian-American Commission for Educational Exchange, with additional support by the DKOP-23 Doctoral Excellence Program of the Ministry for Culture and Innovation of Hungary from the source of the National Research, Development and Innovation Fund.
V.\ K.\ Bandaru and P.\ Alves were supported by the Gordon \& Betty Moore Foundation and the Open Quantum Initiative.
E.\ Brann is a participant with the Oak Ridge Institute for Science and Education (ORISE) and was further supported by the National Science Foundation Graduate Research Fellowship Program under Grant No.\ 2137424.
M.\ Losert was supported by a fellowship from the Army Research Office under Award No.\ W911NF-22-1-0090. 
The architecture development reported in this work was supported through the National Science Foundation QLCI-HQAN (Award No.\ 2016136). 
Any opinions, findings, and conclusions or recommendations expressed in this material are those of the author(s) and do not necessarily reflect the views of the National Science Foundation. 
The authors also thank HRL Laboratories for support. 
This research was also sponsored in part by the Army Research Office under Award No.\ W911NF-23-1-0110. 
The views, conclusions, and recommendations contained in this document are those of the authors and are not necessarily endorsed by nor should they be interpreted as representing the official policies, either expressed or implied, of the Army Research Office or the U.S.\ Government. 
The U.S. Government is authorized to reproduce and distribute reprints for U.S.\ Government purposes notwithstanding any copyright notation herein. 
Some of this work was performed using the computing resources and assistance of the UW-Madison Center For High Throughput Computing (CHTC) in the Department of Computer Sciences. 
The CHTC is supported by UW-Madison, the Advanced Computing Initiative, the Wisconsin Alumni Research Foundation, the Wisconsin Institutes for Discovery, and the National Science Foundation, and is an active member of the OSG Consortium, which is supported by the National Science Foundation and the U.S.\ Department of Energy’s Office of Science.

\section*{Data availability}
All data in this work is generated through simulations and the corresponding codes are publicly available (see Code Availability statement below).

\section*{Code availability}
The source code for reproducing the simulations and figures can be accessed at https://zenodo.org/records/18225182~\cite{zenodo}.

\section*{Competing interests}
V.K.B., M.A.E., M.P.L., and M.F.\ declare a related patent application that proposes a scheme for 2D shuttling: US Patent Application No.~18/975615 (currently under review). 
The remaining authors declare no competing interests.

\appendix
\section{Generating random disorder landscapes}
\label{app:landscapes}

\red{In the theoretical models of shuttling used in this work, we consider basis states that label the pockets, in a moving reference frame. 
In this basis, the Hamiltonian parameters $\varepsilon_j$, $t_p^j$, and $\Delta_j$ vary in time as the system traverses a disordered landscape.
Below, we treat these fluctuations statistically, via their spatial covariance functions.}

We use the methods described in~Ref.~\cite{Losert:2024p040322} to generate random spatial disorder landscapes.
The real and imaginary parts of the valley couplings $\Delta$ ($R$ and $I$) are generated independently for the spatial covariance function 
\begin{equation}
    \text{Cov}[\Delta_{R(I)}, \Delta'_{R(I)}] = \frac{\sigma_\Delta^2}{2}\exp \left(-\frac{d^2}{2 l_\text{dot}^2} \right) ,
\end{equation}
where $\Delta$ and $\Delta'$ are the intervalley couplings for dots separated by a distance $d$ across the heterostructure, $l_\text{dot} = \sqrt{\hbar^2 / m_t E_\text{orb}}$ is the dot radius for both dots, $m_t = 0.19 m_e$ is the electron transverse effective mass, and $E_\text{orb}$ is the characteristic orbital energy splitting of an isotropic harmonic oscillator in a parabolic confinement potential. 
In the simulations, we take $\sigma_\Delta = 56.4$~\SI{}{\micro\electronvolt}, as mentioned in the main text.

As discussed in the main text, we also consider the potential energy fluctuations $\delta\varepsilon_{L(R)}$, including contributions from alloy disorder, lever-arm fluctuations, and charge offsets.
We first consider alloy-disorder contributions, which modify the ground-state energy of the moving pocket and can be modeled as
\begin{equation}
    \delta \varepsilon_\text{alloy} = \int d^3r \; U_\text{qw}(\mathbf{r}) |\psi_\text{env}(\mathbf{r})|^2,
\end{equation}
where $U_\text{qw}$ is the locally varying quantum well confinement potential, including the effects of random-alloy disorder, and $\psi_\text{env}$ is the quantum dot envelope function.
Similar to the intervalley coupling, this quantity is described by the Gaussian spatial covariance function 
\begin{equation}
    \text{Cov}[\delta \varepsilon_\text{alloy}, \delta \varepsilon_\text{alloy}'] = \sigma_\Delta^2 \exp \left(-\frac{d^2}{2 l_\text{dot}^2} \right).
\end{equation}
(More details to be provided in a forthcoming publication.)

We adopt a heuristic approach for the remaining fluctuating parameters in our simulations.
For geometry-induced potential fluctuations of a given device $\delta \varepsilon_\text{gate}$, we assume a Gaussian distribution of fluctuations with zero mean and a spatial covariance characterized by the gate pitch $P$ as
\begin{equation}
    \text{Cov}[\delta \varepsilon_\text{gate}, \delta \varepsilon_\text{gate}'] = \sigma_{\varepsilon,\text{gate}}^2 \exp \left(-\frac{d^2}{2 P^2} \right).
\end{equation} 
From typical experimental values of the lever-arm fluctuations~\cite{Neyens:2024p80, DeSmet:2025p866}, we infer that the standard deviation of such potential fluctuations is of order $\sigma_{\varepsilon,\text{gate}} \approx 1$~\SI{}{\milli\electronvolt}.

For our master equation simulations of 2D shuttling geometries, we also include fluctuations of the tunnel couplings between dots. 
Since such tunnel couplings are real, positive quantities, we assume that they follow a log-normal distribution with a mean value of $t_0$ and a standard deviation of $\sigma_t$. 
The spatial covariance is again described by the gate pitch:
\begin{equation}
    \text{Cov}\left[\ln\left(\frac{t_p}{t_0}\right), \ln\left(\frac{t_p'}{t_0}\right)\right] = \ln\left(1 + \frac{\sigma_t^2}{t_0^2}\right) \exp \left(-\frac{d^2}{2 P^2} \right).
    \label{eq:tpcov}
\end{equation} 
Based on the results of our electrostatics simulations, described below, the tunnel couplings used for the master equation fall in the range of $10^{-6}~\SI{}{\micro\electronvolt} \le t_0 \le 1~\SI{}{\micro\electronvolt}$, while $\sigma_t \approx t_0/10$, which we use in our simulations.

\section{Schr{\"o}dinger-Poisson simulations}
\label{app:electrostatics}

We use the MaSQE Schr{\"o}dinger-Poisson (SP) software package to simulate both multichannel and 2D shuttling devices \cite{Anderson:2022p065123}.
In the SP simulation module, only confinement in the $z$-direction is treated quantum mechanically.
The heterostructure is modeled as a $10$~\SI{}{\nano\meter} Si quantum well sandwiched between a thick $\text{Si}_{0.67}\text{Ge}_{0.33}$ effective substrate and a $40$~\SI{}{\nano\meter} $\text{Si}_{0.67}\text{Ge}_{0.33}$ spacer layer, with metal finger gates on the top of the stack.
Since the pockets are nominally identical, the energy detuning between the pockets $\varepsilon$ is defined as the energy difference between the potential minima of neighboring pockets.

To determine the tunnel couplings between neighboring pockets, we consider a standard two-level double-dot Hamiltonian 
\begin{equation} \label{eq:ham_2level}
H_{2 \times 2} = \frac{\varepsilon}{2} \tau_z + t \tau_x,
\end{equation}
where $t$ equals $t_c$ for the multichannel case and $t_p$ for the 2D case.
Diagonalizing Eq.~(\ref{eq:ham_2level}) identifies the minimum energy gap between the ground and excited states as $2 t$.
This quantity is determined from simulations as follows.
We first perform SP simulations of realistic device geometries and extract the 2D potential energy of the quantum well at the $z$ coordinate corresponding to the center of the quantum well.
We then import this potential into a discretized 2D Schr{\"o}dinger equation and solve to obtain the ground and first excited orbital-state energies.
The simulations are repeated many times while varying the in-plane electric field along the axis between the potential pockets, representing the detuning parameter.
Finally, we identify the minimum energy gap as $2t$.

\section{Time-evolution simulations}
\label{app:time-evolution}

For the time-evolution simulations of the multichannel device, we use the QuTiP Python framework \cite{Johansson:2012p1760}.
Some simulations are performed at the Center for High Throughput Computing at UW-Madison \cite{chtc}.

For the time-evolution simulations of the 2D shuttler, we develop our own numerical methods, based on the theoretical approach described below. 
We first determine the static Hamiltonian at each time step, making use of the spatially varying random potential landscapes described above, including fluctuating intervalley couplings, detuning parameters, and tunnel couplings.
We then perform a unitary transformation into the instantaneous energy eigenbasis, $\{|n\rangle\}$, where the addition of decoherent terms in the Lindblad Eq.~\eqref{eq:Lindblad} are straightforward to compute, once the relaxation rates are known. 
We note that the instantaneous eigenbasis is 10-dimensional, like the original localized basis $\{|z_s,d_j\rangle\}$.

To compute the relaxation rates, we focus on phonon-mediated relaxation processes, making use of derivations given in Refs.~\cite{Krzywda:2021p075439,Hosseinkhani:2021p085309}.
We briefly outline this approach as follows, with a more detailed description given in Appendix~\ref{app:relaxation}.
We consider transitions within the instantaneous eigenbasis, from state $|n\rangle$ to $|m\rangle$, with eigenvalues $E_n$ and $E_m$. 
The corresponding Lindblad collapse operator is given by $L_{nm} = |m\rangle\langle n|$, with a relaxation rate given by
\begin{equation}
    \begin{aligned}
    \Gamma_{nm} =~&\frac{1}{\hbar^2} S_\text{os}(\omega_{nm}) \sum_{j=1}^5 \left|\langle m | {K}_\text{os}^j | n\rangle\right|^2 \\
    &+ \frac{1}{\hbar^2} S_\text{hop} (\omega_{nm}) \sum_{j=2}^5 \left|\langle m | {K}_\text{hop}^j | n\rangle\right|^2 .
    \end{aligned}  
    \label{eq:Gammanm}
\end{equation} 
As shown in~\cite{Krzywda:2021p075439,Hosseinkhani:2021p085309} and Appendix~\ref{app:relaxation}, this expression can be derived from Fermi's golden rule, where we define {$\hbar\omega_{nm} = E_n - E_m$.
Here ${K}_\text{os}$ and ${K}_\text{hop}$ describe the on-site and hopping transitions between pockets:
\begin{gather}
    {K}_\text{os}^j = {\sum_{s=\pm}} | z_s, d_j \rangle \langle z_s, d_j | , \label{eq:coupling} \\
    {K}_\text{hop}^j = {\sum_{s=\pm}} | z_s, d_1 \rangle \langle z_s, d_j | + \mathrm{h.c.} ,
    \label{eq:coupling2}
\end{gather} 
where we make use of the notations defined in Eqs.~\eqref{eq:onsite}-\eqref{eq:valcoup}. 

The `on-site' prefactor in Eq.~(\ref{eq:Gammanm}), given by
\begin{equation}
    S_\text{os}(\omega) = \frac{{\hbar}\omega^3}{8\pi^2\varrho} \left[ \frac{\Xi_d^2 I^l_0 + 2\Xi_d\Xi_u I^l_2 + \Xi_u^2 I^l_4}{c_l^5} + \frac{\Xi_u^2 I^t}{c_t^5}\right] ,
    \label{eq:spectral_density}
\end{equation} 
includes the phonon density of states and the on-site phonon matrix elements, defined as
\begin{gather}
        I^l_k = \int\displaylimits_0^\pi\mathrm{d}\vartheta \int\displaylimits_0^{2\pi}\mathrm{d}\varphi\,\sin\vartheta\cos^k\vartheta\,|\langle d_1|e^{i\mathbf{q}_l\cdot\mathbf{r}}|d_1\rangle|^2 , 
        \label{eq:integrals} \\
        I^t = \int\displaylimits_0^\pi\mathrm{d}\vartheta \int\displaylimits_0^{2\pi}\mathrm{d}\varphi\,\sin^3\vartheta\cos^2\vartheta\,|\langle d_1|e^{i\mathbf{q}_t\cdot\mathbf{r}}|d_1\rangle|^2 .
    \label{eq:integrals2}
\end{gather} 
In these equations, we define the mass density of silicon, $\varrho = 2330~\text{kg/m}^3$, the longitudinal and transversal speeds of sound, $c_l = 9330~\text{m/s}$ and $c_t = 5420~\text{m/s}$, and the deformation potentials $\Xi_d = 5~\SI{}{\electronvolt}$ and $\Xi_u = 8.77~\SI{}{\electronvolt}$. 
The longitudinal and transversal phonon wave vectors take the form $\mathbf{q}_{l,t} = (\omega/c_{l,t}) (\sin\vartheta\cos\varphi,\sin\vartheta\sin\varphi,\cos\vartheta)$. 

The `hopping' prefactor $S_\text{hop}$ in Eq.~(\ref{eq:Gammanm}), describing relaxation from the central pocket to its neighbors, is computed similarly.
However, in this case, the matrix element appearing in the $I^l_k$ and $I^t$ integrals should be replaced by $\langle d_1| e^{i\mathbf{q}_{l,t}\cdot\mathbf{r}} |d_2\rangle$.
For the gaussian pocket wavefunctions described above, we find $S_\text{hop}(\omega) \approx S_\text{os}(\omega) {\exp(-8P^2/l_\text{dot}^2)}$. 

The results described above pertain only to phonon-mediated relaxation processes. 
While other relaxation mechanisms may be present (e.g. charge-noise-mediated effects), it has been shown that phonon mechanisms are dominant at energy scales of 0.1-1~$\SI{}{\milli\electronvolt}$ (corresponding to a magnetic field range of 1-10~T in~\cite{Huang:2014p195302}).
We therefore neglect these other effects here. 
For device parameters $P = 50~\SI{}{\nano\meter}$ and $W = 5~\SI{}{\nano\meter}$, and tunnel couplings in the range of $1~\SI{}{\pico\electronvolt} \le t_0 \le 1~\SI{}{\micro\electronvolt}$, we estimate a wide range of relaxation rates $10^{-12}~\SI{}{\second}^{-1} \le \Gamma_{nm} \le 1~\SI{}{\second}^{-1}$, for the $\Gamma_{nm}$ values that are nonzero.

We then solve the time evolution using a matrix exponentiation technique, separating diagonal and off-diagonal components of the density operator for higher numerical efficiency. 
In detail, we perform a vectorization or flattening of the density operator, representing it as a $100$-component vector instead of a $10\times 10$ matrix through the mapping $|n\rangle\langle m| \mapsto |n\rangle\otimes |m\rangle \equiv |nm\rangle$. 
For any operators $A$ and $B$, their action is mapped as $A|n\rangle\langle m|B \mapsto A|n\rangle \otimes B^T|m\rangle \equiv (A \otimes B^T)|nm\rangle$.
Note that such transformations are basis-dependent; in this case, we specifically choose the energy eigenbasis.
In this representation, the Lindbladian superoperator $\mathcal{L}$ defined by the right-hand side of Eq.~\eqref{eq:Lindblad} can be conveniently expressed as follows. 
For the energy eigenbasis, the Hamiltonian is given by
\begin{equation}
    H = \sum_{n=1}^{10} E_n |n\rangle\langle n | ,
\end{equation} 
and the collapse operators are defined as
\begin{equation}
    L_{nm} = |m\rangle\langle n| .
\end{equation}
We next rearrange the product operators $H\otimes I$ and $I\otimes H^T$ appearing in the commutator term of the master equation, and the operators $L_{nm}\otimes L_{nm}^*$, $L_{nm}^\dagger L_{nm}\otimes I$, and $I\otimes L_{nm}^T L_{nm}^*$ describing the decoherence, as follows. 
(Note that the star here denotes element-wise complex conjugation, i.e., the adjoint of the transpose.) 
After a straightforward derivation, the Lindbladian superoperator is found to be a sum $\mathcal{L} = \mathcal{D} + \mathcal{O}$ of the terms
\begin{equation}
    \begin{aligned}
        \mathcal{D} =~& \sum_{n = 1}^{10} \sum_{m = 1}^{10} \Gamma_{nm} \left(|mm\rangle - |nn\rangle\right) \langle nn| , \\
        \mathcal{O} =~& \frac{i}{\hbar}\underset{n \neq m}{\sum_{n = 1}^{10} \sum_{m = 1}^{10}} \left(E_m - E_n\right) |nm\rangle\langle nm| \\ &- \frac12 \underset{n \neq m}{\sum_{n = 1}^{10} \sum_{m = 1}^{10}} \sum_{\ell = 1}^{10} \left(\Gamma_{n\ell} + \Gamma_{m\ell}\right) |nm\rangle\langle nm| .
    \end{aligned}
\end{equation} 
Here, the first (second) term affects only the diagonal (off-diagonal) components of the density operator. 
Importantly, $\mathcal{O}$ is diagonal in the direct-product (flattened) representation. 
In this way, we can separately calculate the time evolution of the diagonal components by exponentiating $\mathcal{D}$, and that of the off-diagonal components by exponentiating the matrix elements of $\mathcal{O}$. 
Finally, we perform the inverse of the initial unitary transformation, to return to the localized basis. 

Repeating this procedure at every time step, we obtain the probability of leakage as a function of time, as depicted in Fig.~\ref{fig:3}(d). 
Note that obtaining accurate time evolutions requires that the simulation time steps be chosen such that the corresponding displacement is much smaller than the correlation lengths of the random landscapes, $l_\text{dot}$ and $P$.

\section{Relaxation rate calculations}
\label{app:relaxation}

In the following, we provide a more detailed derivation of the relaxation rate formula, Eq.~\eqref{eq:Gammanm} in Appendix~\ref{app:time-evolution}. 
First, we recall that the electron-phonon interaction in the deformation potential approximation takes the form of Eq.~(43) in Ref.~\cite{Hosseinkhani:2021p085309}:
\begin{equation}
    H_\mathrm{e-ph} = H_\mathrm{e-ph}^l + H_\mathrm{e-ph}^t,
\end{equation} where the two terms correspond to longitudinal and transversal phonons, respectively. 
The full expressions are given by
\begin{equation}
    \begin{aligned}
        H_\mathrm{e-ph}^l = i \sum_{\mathbf{q}_l} &\sqrt{\frac{\hbar q_l}{2\varrho V c_l}} \left(\Xi_d + \Xi_u\cos^2\vartheta\right)  \\
        &\times\left(b_{\mathbf{q}_l} - b_{-{\mathbf{q}_l}}^\dagger\right) e^{i{\mathbf{q}_l}\cdot\mathbf{r}} ,
        \end{aligned}
\end{equation}
\begin{equation}
    \begin{aligned}
        H_\mathrm{e-ph}^t = i \Xi_u \sum_{\mathbf{q}_t} &\sqrt{\frac{\hbar q_t}{2\varrho V c_t}} \cos\vartheta\sin\vartheta  \\
        &\times\left(b_{\mathbf{q}_t} + b_{-{\mathbf{q}_t}}^\dagger\right) e^{i{\mathbf{q}_t}\cdot\mathbf{r}} ,
        \end{aligned}
\end{equation} 
where $V$ is the volume of the system, $b_{\mathbf{q}_l}$ and $b_{\mathbf{q}_t}$ are annihilation operators for longitudinal and transversal phonons with wave vectors $\mathbf{q}_l$ and $\mathbf{q}_t$, respectively, and $q_{l,t}=|\mathbf{q}_{l,t}|$. 
All other notations are defined below Eq.~\eqref{eq:integrals2} of Appendix~\ref{app:time-evolution}.

To obtain the relaxation rate, we use Fermi's golden rule to study transitions between the instantaneous energy eigenstates $|n\rangle$ and $|m\rangle$, where $\{|n\rangle\}$ is the 10-dimensional basis set of valley-pocket states described in the main text.
To satisfy energy conservation, this transition should be accompanied by emission of a phonon of energy $\hbar\omega_{nm} = E_n - E_m$, and wave vector $q_{l,t} = \omega_{nm} / c_{l,t}$. 
Taking into account both the longitudinal and transversal phonon modes, with densities of state $g_l(E)$ and $g_t(E)$, we arrive at
\begin{equation}
    \Gamma_{nm} = \Gamma_{nm}^l + \Gamma_{nm}^t ,
\end{equation} where 
\begin{equation}
    \begin{aligned}
    \Gamma_{nm}^{l,t} = \frac{2\pi}{\hbar} \left|\langle m,1_{\mathbf{q}_{l,t}} | H_\mathrm{e-ph}^{l,t} | n,0 \rangle\right|^2 g_{l,t}(E_n - E_m) .
    \end{aligned}
    \label{eq:Glt}
\end{equation} 
Here, $|0 \rangle$ represents the phonon vacuum state, while $|1_{\mathbf{q}_{l,t}} \rangle$ denotes a longitudinal or transversal single-phonon state of wave vector $\mathbf{q}_{l,t}$. 
Due to the typical low temperatures of shuttling devices, we may neglect any thermal excitations of phonons. 
The densities of states may be replaced by a summation over phonon modes of energy $\hbar\omega_{nm}$. 
In realistic crystals, with a macroscopic number of atoms, this summation can be replaced by an integral over a sphere of radius $q_{l,t}$ in reciprocal space:
\begin{equation}
    \Gamma_{nm}^{l,t} = \frac{V q_{l,t}^2}{4\pi^2\hbar^2 c_{l,t}} \int\mathrm{d}\Omega \,\left|\langle m,1_{\mathbf{q}_{l,t}} | H_\mathrm{e-ph}^{l,t} | n,0 \rangle\right|^2  .
    \label{eq:GammaS5}
\end{equation} 
Here, the integral over the solid angle $\Omega$ can be expressed in terms of the polar angle $\vartheta$ and azimuthal angle $\varphi$, as consistent with our notation in Eqs.~\eqref{eq:integrals} and \eqref{eq:integrals2}:
\begin{equation}
    \int\mathrm{d}\Omega \sim \int\displaylimits_0^\pi\mathrm{d}\vartheta\,\sin\vartheta \int\displaylimits_0^{2\pi}\mathrm{d}\varphi  .
\end{equation}
Since there are no phonons in the vacuum state of the matrix element in Eq.~(\ref{eq:GammaS5}), there is no contribution from the annihilation operators, and the relaxation rates reduce to
\begin{gather}
        \Gamma_{nm}^{l} = \frac{\omega_{nm}^3}{8\pi^2\varrho\hbar c_{l}^5}\int\mathrm{d}\Omega \, \left(\Xi_d + \Xi_u\cos^2\vartheta\right)^2 \left|\langle m| e^{i\mathbf{q}_{l}\cdot\mathbf{r}} | n\rangle\right|^2 ,
        \label{eq:Gammanm_l} \\
        \Gamma_{nm}^{t} = \frac{\omega_{nm}^3}{8\pi^2\varrho\hbar c_{t}^5}\int\mathrm{d}\Omega \, \left(\Xi_u \cos\vartheta \sin\vartheta\right)^2 \left|\langle m| e^{i\mathbf{q}_{t}\cdot\mathbf{r}} | n\rangle\right|^2 .
\label{eq:Gammanm_t}
\end{gather} 

To evaluate the matrix elements $\langle m| e^{i\mathbf{q}_{l,t}\cdot\mathbf{r}} | n\rangle$, we expand $| n\rangle$ and $| m\rangle$ in terms of the localized basis states $|z_s,d_j\rangle$, described in the main text:
\begin{equation}
    \begin{aligned}
    |n\rangle = \sum_{s = \pm} \sum_{j = 1}^5 \nu_{s,j} |z_s,d_j\rangle ,\\
    |m\rangle = \sum_{s = \pm} \sum_{j = 1}^5 \mu_{s,j} |z_s,d_j\rangle .
    \end{aligned}
\end{equation} 
Now, using the fact that $\langle z_+|e^{i\mathbf{q}_{l,t}\cdot\mathbf{r}}|z_-\rangle\approx 0$, due to the very different magnitudes of valley and phonon wavevectors ($k_0\gg q_{l,t}$), the phonon matrix elements reduce to
\begin{equation}
    \langle m| e^{i\mathbf{q}_{l,t}\cdot\mathbf{r}} | n\rangle \approx \sum_{s = \pm} \sum_{j,k = 1}^5 \mu_{s,j}^* \nu_{s,k} \langle d_j| e^{i\mathbf{q}_{l,t}\cdot\mathbf{r}} | d_k\rangle .
\end{equation} 
Since the states $|d_j\rangle$ are strongly localized at the pockets, the matrix elements $\langle d_j| e^{i\mathbf{q}_{l,t}\cdot\mathbf{r}} | d_k\rangle$ have a significant amplitude only when $j=k$, or when $j$ and $k$ correspond to nearest neighbors, such that
\begin{equation}
    \begin{aligned}
    \langle m| e^{i\mathbf{q}_{l,t}\cdot\mathbf{r}} | n\rangle \approx \sum_{s = \pm} \sum_{j = 1}^5 \mu_{s,j}^* \nu_{s,j} \langle d_j| e^{i\mathbf{q}_{l,t}\cdot\mathbf{r}} | d_j\rangle  \\ + \sum_{s = \pm} \sum_{j = 2}^5 \left[\mu_{s,1}^* \nu_{s,j} \langle d_1| e^{i\mathbf{q}_{l,t}\cdot\mathbf{r}} | d_j\rangle + \mathrm{h.c.} \right] .
    \end{aligned}
    \label{eq:matrix_element}
\end{equation} Next, we observe that, since the pockets are nearly identical in shape and size, the magnitudes of the matrix elements $\langle d_j| e^{i\mathbf{q}_{l,t}\cdot\mathbf{r}} | d_j\rangle$ and $\langle d_1| e^{i\mathbf{q}_{l,t}\cdot\mathbf{r}} | d_j\rangle$ do not depend $j$. 
We may therefore transform between pocket indices in these integrals by performing a change of variables, $\mathbf{r} \to \mathbf{r} - \mathbf{r}_j$, such that
\begin{equation}
    \begin{aligned}
    \langle d_j| e^{i\mathbf{q}_{l,t}\cdot\mathbf{r}} | d_j\rangle = e^{i\mathbf{q}_{l,t}\cdot(\mathbf{r}_j - \mathbf{r}_1)} \langle d_1| e^{i\mathbf{q}_{l,t}\cdot\mathbf{r}} | d_1\rangle , \\
    \langle d_1| e^{i\mathbf{q}_{l,t}\cdot\mathbf{r}} | d_j\rangle = e^{i\mathbf{q}_{l,t}\cdot(\mathbf{r}_j - \mathbf{r}_2)} \langle d_1| e^{i\mathbf{q}_{l,t}\cdot\mathbf{r}} | d_2\rangle .
    \end{aligned}
\end{equation} 

The relaxation rates in Eqs.~\eqref{eq:Gammanm_l} and \eqref{eq:Gammanm_t} require squaring the absolute value of Eq.~\eqref{eq:matrix_element}, yielding two types of product terms.
As explained below, the self-product terms are found to dominate in the relaxation rate integrals, while the cross-product terms have a much smaller contribution.
This is due to the complex phases in the latter terms, which cause oscillations that suppress the rate integrals, for typical dot parameters.
We therefore remove the cross-product terms here (see below), obtaining
\begin{equation}
    \begin{aligned}
    \left|\langle m | e^{i\mathbf{q}_{l,t}\cdot\mathbf{r}} | n\rangle\right|^2 \rightarrow \sum_{j = 1}^5 \left| \sum_{s = \pm} \mu_{s,j}^* \nu_{s,j} \right|^2 \left| \langle d_1| e^{i\mathbf{q}_{l,t}\cdot\mathbf{r}} | d_1\rangle \right|^2 \\ + \sum_{j = 2}^5 \left| \sum_{s = \pm} \mu_{s,1}^* \nu_{s,j} + \mathrm{c.c.}\right|^2 \left|\langle d_1| e^{i\mathbf{q}_{l,t}\cdot\mathbf{r}} | d_2\rangle\right|^2 .
    \end{aligned} 
    \label{eq:square}
\end{equation} 

Here, the $s$ sums contain only expansion coefficients of the energy eigenstates; they can be written more compactly as
\begin{gather}
        \left| \sum_{s = \pm} \mu_{s,j}^* \nu_{s,j} \right|^2 = \left|\langle m | {K}_\text{os}^j | n\rangle\right|^2 , \\
        \left| \sum_{s = \pm} \mu_{s,1}^* \nu_{s,j} + \mathrm{c.c.}\right|^2 = \left|\langle m | {K}_\text{hop}^j | n\rangle\right|^2 ,
\end{gather} 
where we introduce the operators
\begin{gather}
    {K}_\text{os}^j = {\sum_{s=\pm}} | z_s, d_j \rangle \langle z_s, d_j | , \label{eq:coupling_suppl} \\
    {K}_\text{hop}^j = {\sum_{s=\pm}} | z_s, d_1 \rangle \langle z_s, d_j | + \mathrm{h.c.} 
    \label{eq:coupling_suppl2}
\end{gather} 
The angular integrals of Eqs.~\eqref{eq:Gammanm_l} and \eqref{eq:Gammanm_t} therefore act only on the matrix elements, giving~\cite{Hosseinkhani:2021p085309}
\begin{equation}
    S_\text{os}(\omega) = \frac{{\hbar}\omega^3}{8\pi^2\varrho} \left[ \frac{\Xi_d^2 I^l_0 + 2\Xi_d\Xi_u I^l_2 + \Xi_u^2 I^l_4}{c_l^5} + \frac{\Xi_u^2 I^t}{c_t^5}\right] ,
    \label{eq:spectral_density_suppl}
\end{equation} 
for the $|\langle d_1| e^{i\mathbf{q}_{l,t}\cdot\mathbf{r}} | d_1\rangle|^2$ term,
where we define
\begin{gather}
        I^l_k = \int\displaylimits_0^\pi\mathrm{d}\vartheta \int\displaylimits_0^{2\pi}\mathrm{d}\varphi\,\sin\vartheta\cos^k\vartheta\,|\langle d_1|e^{i\mathbf{q}_l\cdot\mathbf{r}}|d_1\rangle|^2 , 
        \label{eq:integrals_suppl} \\
        I^t = \int\displaylimits_0^\pi\mathrm{d}\vartheta \int\displaylimits_0^{2\pi}\mathrm{d}\varphi\,\sin^3\vartheta\cos^2\vartheta\,|\langle d_1|e^{i\mathbf{q}_t\cdot\mathbf{r}}|d_1\rangle|^2 ,
    \label{eq:integrals_suppl2}
\end{gather}
and
\begin{equation}
    S_\text{hop}(\omega) = \frac{{\hbar}\omega^3}{8\pi^2\varrho} \left[ \frac{\Xi_d^2 {\tilde I}^l_0 + 2\Xi_d\Xi_u {\tilde I}^l_2 + \Xi_u^2 {\tilde I}^l_4}{c_l^5} + \frac{\Xi_u^2 {\tilde I}^t}{c_t^5}\right] ,
    \label{eq:spectral_density_suppl}
\end{equation} 
for the $|\langle d_1| e^{i\mathbf{q}_{l,t}\cdot\mathbf{r}} | d_2\rangle|^2$ term,
where we define
\begin{gather}
        {\tilde I}^l_k = \int\displaylimits_0^\pi\mathrm{d}\vartheta \int\displaylimits_0^{2\pi}\mathrm{d}\varphi\,\sin\vartheta\cos^k\vartheta\,|\langle d_1|e^{i\mathbf{q}_l\cdot\mathbf{r}}|d_2\rangle|^2 , 
        \label{eq:integrals_suppl_tilde} \\
        {\tilde I}^t = \int\displaylimits_0^\pi\mathrm{d}\vartheta \int\displaylimits_0^{2\pi}\mathrm{d}\varphi\,\sin^3\vartheta\cos^2\vartheta\,|\langle d_1|e^{i\mathbf{q}_t\cdot\mathbf{r}}|d_2\rangle|^2 .
    \label{eq:integrals_suppl_tilde2}
\end{gather}
Substituting these results into Eqs.~\eqref{eq:Gammanm_l} and \eqref{eq:Gammanm_t}, we finally recover Eq.~\eqref{eq:Gammanm} of Appendix~\ref{app:time-evolution}:
\begin{equation}
    \begin{aligned}
    \Gamma_{nm} =~&\frac{1}{\hbar^2} S_\text{os}(\omega_{nm}) \sum_{j=1}^5 \left|\langle m | {K}_\text{os}^j | n\rangle\right|^2 \\
    &+ \frac{1}{\hbar^2} S_\text{hop} (\omega_{nm}) \sum_{j=2}^5 \left|\langle m | {K}_\text{hop}^j | n\rangle\right|^2 .
    \end{aligned}
\end{equation} 

We conclude this section by providing a brief justification for neglecting the cross-product terms in Eq.~\eqref{eq:square}.
We do this by numerically computing the integrals in Eqs.~\eqref{eq:integrals_suppl}, \eqref{eq:integrals_suppl2}, \eqref{eq:integrals_suppl_tilde}, and \eqref{eq:integrals_suppl_tilde2} for typical dot parameters.
We model the localized basis states $|d_j\rangle$ in the harmonic approximation as Gaussian wave functions:
\begin{equation}
    \langle\mathbf{r}|d_j\rangle = \frac{\delta(z)}{\sqrt{\pi l_\mathrm{dot}^2}} \exp\left[-\frac{(x - x_j)^2 + (y - y_j)^2}{2 l_\mathrm{dot}^2}\right] .
\end{equation} 
For simplicity, we have assumed that confinement in the $z$ direction is much stronger than in the $x$-$y$ directions, approximating it as a Dirac delta function. 
The matrix element $\langle d_1|e^{i\mathbf{q}\cdot\mathbf{r}}|d_1\rangle$ can then be viewed as the Fourier transform of a Gaussian function, such that
\begin{equation}
    \left|\langle d_1|e^{i\mathbf{q}\cdot\mathbf{r}}|d_1\rangle\right|^2 = \exp\left(-q^2 l_\mathrm{dot}^2\sin^2\vartheta/2\right) .
    \label{eq:Fourier_Gauss}
\end{equation} 
With the change of variables $\mathbf{r} \to \mathbf{r} - (\mathbf{r}_1 + \mathbf{r}_2)/2$, the matrix element $\langle d_1|e^{i\mathbf{q}\cdot\mathbf{r}}|d_2\rangle$ similarly reduces to 
\begin{equation}
    \left|\langle d_1|e^{i\mathbf{q}\cdot\mathbf{r}}|d_2\rangle\right|^2 = \exp\left(-\frac{8P^2}{l_\mathrm{dot}^2}\right) \left|\langle d_1|e^{i\mathbf{q}\cdot\mathbf{r}}|d_1\rangle\right|^2 .
\end{equation} 
This shows that the spectral densities corresponding to the on-site and hopping terms differ only by a factor of $\exp\left(-8P^2/l_\mathrm{dot}^2\right)$, as mentioned in Appendix~\ref{app:time-evolution}. 
The common factor $\exp\left(-q^2 l_\mathrm{dot}^2\sin^2\vartheta/2\right)$ remains in all integrals of Eqs.~\eqref{eq:integrals_suppl}, \eqref{eq:integrals_suppl2}, \eqref{eq:integrals_suppl_tilde}, and \eqref{eq:integrals_suppl_tilde2}, which we solve numerically using quadrature methods, for both the self-product and cross-product terms.

While Eqs.~\eqref{eq:integrals_suppl}, \eqref{eq:integrals_suppl2}, \eqref{eq:integrals_suppl_tilde}, and \eqref{eq:integrals_suppl_tilde2} explicitly describe the self-product integrals, the cross-product terms are defined very similarly.
However, the latter also include phase factors of the form $e^{i\mathbf{q}\cdot(\mathbf{r}_j - \mathbf{r}_\ell)}$. 
This difference has an important consequence when computing the $\varphi$ integral. 
For the direct terms, where $j=l$, this integral yields a constant factor of $2\pi$, while for the cross-product terms, it yields a Bessel function of the first kind: $2\pi J_0\left(q|\mathbf{r}_j - \mathbf{r}_\ell|\sin\vartheta\right)$. 
The latter is a highly oscillatory function, so that if $q|\mathbf{r}_j - \mathbf{r}_\ell|\gg q l_\mathrm{dot}$, where the latter appears in Eq.~\eqref{eq:Fourier_Gauss}, the $\vartheta$ integral is strongly suppressed. For typical phonon wavelengths of 20-40~$\SI{}{\nano\meter}$, a gate pitch of $P=50~\SI{}{\nano\meter}$, and orbital energy splittings larger than $1.5~\SI{}{\milli\electronvolt}$, the cross-product integrals are then found to be at most $0.5\%$ in magnitude of the direct-product integrals. 
For a slightly larger gate pitch of $70~\SI{}{\nano\meter}$, consistent with several recent experiments, this ratio is even smaller, below $0.25\%$.
Since even the main relaxation effects observed in the simulations of Fig.~\ref{fig:3}(d) are weak, corrections of this order are essentially irrelevant, justifying our previous claim about the size of the cross-product terms.

\bibliography{bibliography}

\end{document}